\documentclass[prd,%
preprint,%
tightenlines,groupedaddress,showpacs,%
nofootinbib,eqsecnum,amsfonts,amsmath,amssymb]{revtex4} \usepackage{bm}
\usepackage{graphicx}

\begin{document}

\title{Gravitational radiation reaction in the equations of motion of compact
binaries to 3.5 post-Newtonian order}

\date{\today}

\author{Samaya Nissanke}\email{nissanke@iap.fr} \author{Luc
Blanchet}\email{blanchet@iap.fr}
\affiliation{${\mathcal{G}}{\mathbb{R}}\varepsilon{\mathbb{C}}{\mathcal{O}}$
-- Gravitation et Cosmologie,\\ Institut d'Astrophysique de Paris,
C.N.R.S.,\\ 98$^{\text{bis}}$ boulevard Arago, 75014 Paris, France}

\date{\today} 

\begin{abstract} 

We compute the radiation reaction force on the orbital motion of compact
binaries to the 3.5 post-Newtonian (3.5PN) approximation, \textit{i.e.} one PN
order beyond the dominant effect. The method is based on a direct PN iteration
of the near-zone metric and equations of motion of an extended isolated
system, using appropriate ``asymptotically matched'' flat-space-time retarded
potentials. The formalism is subsequently applied to binary systems of point
particles, with the help of the Hadamard self-field regularisation. Our result
is the 3.5PN acceleration term in a general harmonic coordinate frame.
Restricting the expression to the centre-of-mass frame, we find perfect
agreement with the result derived in a class of coordinate systems by Iyer and
Will using the energy and angular momentum balance equations.
\end{abstract}

\maketitle

\section{Introduction}\label{secI}

Since the discovery of the Einstein field equations, gravitational radiation
has remained a matter of major theoretical interest, and has led to extensive
theoretical studies on the nature and origin of gravitational wave emission
from isolated sources. Approximation methods in General Relativity, such as
the post-Newtonian (PN) expansion, result in both accurate and measurable
details of the emission. The prospect of the detection of gravitational waves
bathing the Earth by ground and space based interferometers LIGO, VIRGO and
LISA, provides a further impetus to such theoretical investigations. Not only
would such detectors enable an important comparison between astrophysical
observations and theoretical predictions, but they would also ultimately provide
strong tests for General Relativity.

A favourable potential source for the detectors is the radiation-reaction
dominated inspiral and eventual coalescence of two compact objects (neutron
stars or black holes). For such systems, which will undergo hundreds to
thousands (depending on the masses) orbital cycles in the frequency bandwidth
of LIGO and VIRGO, relativistic corrections to the Newtonian order in the
orbital phasing and wave form play a crucial role in preparing the theoretical
templates. Indeed, the detection and analysis of these waves in the detectors
require at least a third post-Newtonian (3PN) correction to both the energy
flux radiated at infinity and in the binary's equations of motion.

The objective of this paper is to compute explicitly the radiation reaction
force in the equations of motion of a compact binary system at the 3.5PN order
$\sim\mathcal{O}(c^{-7})$ in harmonic coordinates (in both a general frame and
the centre-of-mass frame). It is well known that the leading-order radiation
reaction effect occurs at the 2.5PN order $\sim\mathcal{O}(c^{-5})$
\cite{CE70,Bu71,Miller74,Ehl80,Kerlick80a,Kerlick80b,PapaL81,BD84} (see
\textit{e.g.} \cite{Dcargese} for a review). We shall, therefore, compute the
radiation reaction at the 1PN relative order, which corresponds, \textit{via}
appropriate balance equations, to the 1PN corrections in the energy and
angular momentum radiated by the system at future null infinity (relative to
the standard quadrupole formul\ae). The 1PN radiation reaction force is also
responsible for the dominant effect in the loss of linear momentum, widely
referred to as the gravitational radiation ``recoil''.

Up to the 3.5PN order, the conservative terms in the equations of motion are
clearly distinct from the non-conservative, radiation reaction, terms. This
clean separation manifests itself as ``even'' (Newtonian, 1PN, 2PN and 3PN)
and ``odd'' (2.5PN and 3.5PN) orders respectively.\,\footnote{In the present
paper we adopt the terminology that an even (respectively odd) term is one
having an even (odd) power of $1/c$ in front.} The 4PN approximation, however,
contains both some conservative terms, and also, a contribution from the
radiation reaction. The former terms are given by some ``instantaneous''
functionals of the source, whilst the latter is associated with the
gravitational wave tails \cite{BD88}, and is given by a ``hereditary'' type
integral, extending over the past history of the source.

The PN assumption of slow-motion limits the validity of the PN expansion to
the so-called near-zone of the source ($r\ll c\,T$, where $T$ is a typical
period of variation of the source). An important consequence of such a
near-zone limitation is that one cannot incorporate directly the radiation
reaction into the local PN expansion, since the radiation effects depend on
the boundary conditions imposed on the radiation field at infinity ($r\gg a$,
where $a$ is the size of the source), notably the famous no-incoming radiation
condition imposed at past null infinity, $r\rightarrow +\infty$ with
$t+r/c=\mathrm{const}$. At present, two different approaches have been
proposed and implemented for treating the problem.

The first method is based on an asymptotic matching between the PN expansion
valid in the near zone, and the multipolar expansion for the field outside the
source. The matching occurs in the so-called exterior near-zone of the source,
defined by $a < r\ll c\,T$. The asymptotic matching was introduced in this
field by Burke and Thorne \cite{BuTh70,Bu71}. The radiation reaction force has
been derived by matching up to the 3.5PN order for general matter systems
\cite{B97}, and even at the 4PN order, which, as we mentioned above, consists
of the contribution of tails \cite{BD88,BD92}. The most developed treatment
for the exterior multipolar expansion is the so-called MPM expansion, which
combines the multipolar (M) expansion with a post-Minkowskian (PM) scheme
\cite{BD86}. The general solution of the matching equation between the MPM
exterior and PN inner fields has recently been obtained \cite{PB02,BFN04}. In
the present calculation, we shall parametrise the PN metric by some
appropriate ``asymptotically matched'' retarded potentials, which incorporate
the 3.5PN radiation reaction effects, and are introduced in
Refs.~\cite{BFP98,BFeom}. These potentials at 3.5PN order result from some
direct integration of the field equations by means of retarded integrals in
the same manner as in Ref.~\cite{ADec75}. The end result, which we shall
obtain, has already been determined by K{\" o}nigsd{\" o}rffer, Faye and
Sch\"afer \cite{KFS03} (based on the previous works \cite{S85,JaraS97}) within
the framework of the ADM Hamiltonian formalism, and by Pati and Will
\cite{PW02} (see also \cite{WWi96,PW00}) using their variant iteration of the
relaxed Einstein field equations in harmonic coordinates.

The second method is exclusively applicable to compact binary systems
(modelled by point particles). It consists of using the known PN expressions
for the energy and angular momentum radiated at infinity, and of assuming that
these fluxes are balanced by the corresponding losses of energy and angular
momentum in the binary's local equations of motion. This method, based on the
energy and angular momentum balance equations, has been developed by Iyer and
Will \cite{IW93,IW95}. As shown in Refs.~\cite{IW93,IW95}, the requirement of
energy and momentum balance determines uniquely the radiation reaction force
at 3.5PN order in a class of coordinate systems. The residual coordinate
freedom is entirely specified by two arbitrary gauge parameters at 2.5PN order
and by six further ones at 3.5PN. The 2.5PN parameters assume some specific
values in the case of the scalar radiation reaction potential of Burke and
Thorne at 2.5PN order. They are specified by other values in the case of the
2.5PN radiation reaction in harmonic coordinates as calculated by Damour and
Deruelle \cite{DD81a,D82,D83houches}. At 3.5PN order, there is also complete
agreement, in the sense that a unique set of 3.5PN gauge parameters can be
determined each time, with the scalar and vectorial radiation reaction
potentials of Blanchet \cite{B97}, which are valid in some extended
Burke-Thorne type gauge, with the end result of Pati and Will \cite{PW02}, who
work in harmonic coordinates, and with K{\" o}nigsd{\" o}rffer \textit{et al.}
\cite{KFS03}, who use ADM coordinates. The method of balance equations has
also been extended up to 4.5PN order (excluding the tails at 4PN) in
Ref.~\cite{GopuII97}.

The two previous approaches have, therefore, enabled the successful
determination of the 2.5 and 3.5PN radiation reaction terms of the compact
binaries' orbital dynamics. Our new derivation at 3.5PN order is in complete
agreement with the latter works, and, in particular, we confirm the earlier
result of \cite{PW02}. The principles of the method followed in
Ref.~\cite{PW02} are similar to ours, since both methods are based on the PN
iteration of the Einstein field equations relaxed by the harmonic coordinate
condition. There are, however, important differences in the implementation of
the asymptotic matching procedure, as well as several more minor technical
differences. In fact, our own method is justified by the end result of the
particular matching procedure we use, as given in Ref.~\cite{BFN04}.
Furthermore, though of less relevance in the present problem, the external
field in our approach is described by a choice of multipole moments which is
different from the choice adopted in \cite{WWi96,PW00,PW02}. In addition, our
treatment of the compact objects model the particles by delta function
singularities, with the help of the Hadamard self-field regularisation, whilst
Pati and Will \cite{PW02} do not implement a regularisation scheme and
instead, model each of the bodies by some spherical, non-rotating, extended
fluid balls.

The theoretical framework of the present paper is the 3.5PN equations of
motion of a general matter system, derived in \cite{BFeom} by a direct PN
iteration of the metric in harmonic coordinates. As mentioned previously, the
metric is expressed as a functional of a particular set of non-linear retarded
potentials. Such a metric is then specialised to the model of two delta
function singularities by using the standard prescription for a distributional
stress-energy tensor in General Relativity. To cure the divergencies
associated with each particle's infinite self-field, we systematically apply
the Hadamard ``partie-finie'' regularisation \cite{Hadamard,Schwartz,Sellier},
or, more precisely, a specific variant of it defined in \cite{BFreg,BFregM}.
In addition to the divergencies due to the point particles singularities,
which are dealt with by Hadamard's regularisation, Poisson-like integrals
arise at high PN orders and are typically divergent at ``spatial infinity''.
However, this problem is technically overcome by the introduction of
alternative, general solutions to the Poisson equation, in the form of some
regularised versions of the usual Poisson integral, constructed from a
specific finite part procedure called $\mathrm{FP}$. This conforms with our
definition of the multipole moments in the external field (Ref.~\cite{PB02}
describes the general formalism which allows one to use such a finite part).
Furthermore, the recent work \cite{BFN04} shows that the same finite part
$\mathrm{FP}$ is also to be applied when computing the 3.5PN radiation
reaction effects.

When investigating the equations of motion at the 3.5PN level, we will not
encounter any logarithmic divergencies of integrals similar to the ones found
at the 3PN conservative level. Such logarithmic divergencies are responsible
for an ``incompleteness'' of Hadamard's regularisation in treating the
sources' singular nature. Indeed, this incompleteness results in the
appearance of one physical undetermined parameter, called $\lambda$, in the
3PN equations of motion in harmonic coordinates \cite{BF00,BFeom}, or
alternatively, the equivalent parameter known as the static ambiguity
$\omega_s$ in the 3PN Hamiltonian of the particles in ADM-type coordinates
\cite{JaraS98,JaraS99}. A recent application of dimensional regularisation, in
the framework of which the logarithmic divergencies correspond to poles when
the spatial dimension $d$ approaches 3, yielded the numerical value of these
parameters and showed that they are indeed equivalent
\cite{DJSdim,BDE04}.\,\footnote{Other ambiguity parameters, $\xi$, $\kappa$
and $\zeta$, present in the radiation field of point particles binaries at 3PN
order, have also been resolved by means of dimensional regularisation
\cite{BDEI04}.} A complete calculation of the 3PN equations of motion has also
been performed using an independent method by Itoh \textit{et al.}
\cite{itoh1,itoh2}. Here, for the computation of the terms at 3.5PN order, we
can use Hadamard's standard regularisation or any of the proposed variants of
it without encountering such problems. In fact, since there are no logarithmic
divergencies at the 3.5PN order, nor any associated poles, the result obtained
from Hadamard's regularisation is identical to the one arising from
dimensional regularisation.

The plan of this paper is as follows. In Section \ref{secII}, we present the
expressions for the metric and equations of motion up to 3.5PN order for
general isolated matter systems. Section \ref{secIII} considers the specific
mathematical model of two compact objects, which will be described by
point-particle singularities. All the elementary potentials needed for the
3.5PN radiation reaction terms are computed in Section \ref{secIV}. Finally,
we present our final result for the binary's 3.5PN acceleration in Section
\ref{secV}, where we also compare it with the existing literature.

\section{Formalism for general matter systems}\label{secII}

This Section presents the 3.5PN equations of motion, which are expressed as a
function of a particular set of elementary non-linear potentials, and are
valid for a general smooth hydrodynamical ``fluid'' system in harmonic
coordinates. Thus, we assume (initially) that the matter system possesses
neither singularities nor black holes, and can be described by some
Eulerian-type equations involving some high relativistic corrections.

\subsection{Definition of a set of retarded non-linear potentials}\label{secIIa}

We begin by stating the result of the direct PN iterative method for the
near-zone metric, valid for $r\ll c\,T$, which is parametrised by the retarded
potentials (given by some retarded integral) introduced in
Refs.~\cite{BFP98,BFeom}. Convenience and convention dictate the specific form
of these retarded potentials. The near-zone metric asymptotically matches to
an exterior far-zone radiative-type metric in the overlapping exterior
near-zone. The exterior metric is known from a multipolar post-Minkowskian
(MPM) formalism \cite{BD86}. The matching to the PN inner metric yields a
solution which is globally defined over all space-time (in a formal sense of
PN expansions) in the harmonic coordinate system \cite{PB02,BFN04}. The 3.5PN
iterated metric is given by \cite{BFeom},
\allowdisplaybreaks{\begin{subequations}\label{metricg}\begin{eqnarray} g_{00}
& = & -1 + \frac{2}{c^{2}}V - \frac{2}{c^{4}} V^{2} + \frac{8}{c^{6}}
\left(\hat{X} + V_{i} V_{i} + \frac{V^{3}}{6}\right)\nonumber\\ && +
\frac{32}{c^{8}} \left(\hat{T} - \frac{1}{2} V \hat{X} + \hat{R}_{i} V_{i} -
\frac{1}{2} V V_{i} V_{i} - \frac{V^{4}}{48}\right)+
\mathcal{O}\left(\frac{1}{c^{10}}\right),\\ g_{0i} & = & - \frac{4}{c^{3}}
V_{i} - \frac{8}{c^{5}} \hat{R}_{i} - \frac{16}{c^{7}} \left(\hat{Y}_{i} +
\frac{1}{2}\hat{W}_{ij} V_{j} + \frac{1}{2} V^{2} V_{i}\right) +
\mathcal{O}\left(\frac{1}{c^{9}}\right),\\ g_{ij} & = & \delta_{ij} \left[1 +
\frac{2}{c^{2}}V + \frac{2}{c^{4}} V^{2} + \frac{8}{c^{6}} \left(\hat{X} +
V_{k} V_{k} + \frac{V^{3}}{6}\right)\right] \nonumber\\ && +
\frac{4}{c^{4}}\hat{W}_{ij} + \frac{16}{c^{6}} \left( \hat{Z}_{ij} + \frac{1}{2} V
\hat{W}_{ij} - V_{i} V_{j} \right) + \mathcal{O}\left(\frac{1}{c^{8}}\right).
\end{eqnarray}\end{subequations}}\noindent
This metric explicitly involves only some \textit{even} powers of $1/c$.
Indeed, the \textit{odd} terms, notably the 3.5PN terms we are looking for,
are implicitly contained in the definitions of the elementary potentials $V$,
$V_i$, $\hat{W}_{ij}$, $\cdots$, which parametrise the metric, as is shown
below when we perform a PN expansion of the retardation of these potentials.

The matter stress-energy tensor, $T^{\mu\nu}$, is conventionally expressed in
terms of certain mass, current and stress densities, given respectively as,
\begin{subequations}\label{sigma}\begin{eqnarray}
\label{sigmaT1}
\sigma & \equiv & \frac{T^{00} + T^{ii}}{c^2},\\
\label{sigmaT2}
\sigma_{i} &  \equiv  & \frac{T^{0i}}{c},\\
\label{sigmaT3}
\sigma_{ij} &  \equiv  &  T^{ij},
\end{eqnarray}\end{subequations}
(where $T^{ii}\equiv\delta_{ij}T^{ij}$). The potentials may be grouped
depending on the PN order at which they initially appear. For the Newtonian
and 1PN orders, they are given by,
\begin{subequations}\label{pot1}\begin{eqnarray}
V & = & \Box_{\mathcal{R}}^{-1}[-4 \pi G\, \sigma],\label{V}\\ V_{i} &
= & \Box_{\mathcal{R}}^{-1}[-4 \pi G\, \sigma_{i}],
\end{eqnarray}\end{subequations}
where $\Box_{\mathcal{R}}^{-1}$ denotes the usual flat-space-time
d'Alembertian retarded ($\mathcal{R}$) integral. Next, the potentials which
appear at the 2PN order are defined by,
\allowdisplaybreaks{\begin{subequations}\label{pot2}\begin{eqnarray} \hat{X} &
= & \Box_{\mathcal{R}}^{-1}\left[-4 \pi G\, V \sigma_{ii} + \hat{W}_{ij}
\partial^2_{ij} V + 2 V_{i} \partial_{t} \partial_{i} V + V \partial_{t}^{2}
V\right.\nonumber\\ & & \qquad\left. + \frac{3}{2}(\partial_{t} V)^{2} - 2
\partial_{i} V_{j}\partial_{j} V_{i}\right],\\ \hat{R}_{i} & = &
\Box_{\mathcal{R}}^{-1}\left[-4 \pi G\, (V \sigma_{i} - V_{i} \sigma) - 2
\partial_{k} V \partial_{i} V_{k} - \frac{3}{2} \partial_{t} V \partial_{i}
V\right], \\ \hat{W}_{ij} & = & \Box_{{\cal R}}^{-1}\left[-4 \pi G\,
(\sigma_{ij} - \delta_{ij} \sigma_{kk}) - \partial_{i} V \partial_{j}
V\right].\label{Wij}
\end{eqnarray}\end{subequations}}\noindent
Finally, the relevant potentials for the highest established order of 3PN are,
\footnote{When performing the PN iteration for point particles in the context
of the extended Hadamard regularisation \cite{BFreg,BFregM}, there are some
extra contributions to be added to the 3PN potentials, which are due to the
violation of the Leibniz rule for the derivative of a product by the
distributional derivatives, see Eqs.~(3.27) in \cite{BFeom}. These so-called
``Leibniz'' terms arise, however, only at 3PN order and do not contribute to
the present computation at 3.5PN order, where, as previously discussed, the
Hadamard regularisation can be applied without the problems encountered at the
previous 3PN order.}
\allowdisplaybreaks{\begin{subequations}\label{pot3}\begin{eqnarray} \hat{T} &
= & \Box_{\mathcal{R}}^{-1}\ \left[-4 \pi G\, \left(\frac{1}{4} \sigma_{ij}
\hat{W}_{ij} + \frac{1}{2} V^{2} \sigma_{ii} + \sigma V_{i} V_{i}\right) +
\hat{Z}_{ij} \partial^2_{ij} V + \hat{R}_{i} \partial_{t} \partial_{i} V \right.
\nonumber\\ && \qquad - 2 \partial_{i} V_{j} \partial_{j} \hat{R}_{i} -
\partial_{i} V_{j} \partial_{t} \hat{W}_{ij} + V V_{i} \partial_{t}
\partial_{i} V + 2 V_{i} \partial_{j} V_{i}\partial_{j} V + \frac{3}{2} V_{i}
\partial_{t} V \partial_{i} V \nonumber\\ && \qquad \left. + \frac{1}{2} V^{2}
\partial_{t}^{2} V + \frac{3}{2} V (\partial_{t} V)^{2} - \frac{1}{2}
(\partial_{t} V_{i})^{2} \right],\\ \hat{Y}_{i} & = & \Box_{{\cal R}}^{-1}\
\left[-4 \pi G\, \left(- \sigma \hat{R}_{i} - \sigma V V_{i} + \frac{1}{2}
\sigma_{k} \hat{W}_{ik} + \frac{1}{2} \sigma_{ik} V_{k} + \frac{1}{2}
\sigma_{kk} V_{i}\right) + \hat{W}_{kl} \partial_{kl} V_{i} \right.
\nonumber\\ && \qquad - \partial_{t} \hat{W}_{ik} \partial_{k} V +
\partial_{i} \hat{W}_{kl}\partial_{k} V_{l} - \partial_{k} \hat{W}_{il}
\partial_{l} V_{k} - 2 \partial_{k} V \partial_{i} \hat{R}_{k} - \frac{3}{2}
V_{k} \partial_{i} V \partial_{k} V \nonumber\\ && \qquad \left. - \frac{3}{2}
V \partial_{t} V \partial_{i} V - 2 V \partial_{k} V \partial_{k} V_{i} + V
\partial_{t}^{2} V_{i} + 2 V_{k} \partial_{k} \partial_{t} V_{i} \right], \\
\hat{Z}_{ij} & = & \Box_{\mathcal{R}}^{-1}\ \left[-4 \pi G\, V
\left(\sigma_{ij} - \delta_{ij} \sigma_{kk}\right) - 2 \partial_{(i} V
\partial_{t} V_{j)} + \partial_{i} V_{k} \partial_{j} V_{k} + \partial_{k}
V_{i}\partial_{k} V_{j} \right. \nonumber\\ && \qquad \left. - 2 \partial_{(i}
V_{k} \partial_{k} V_{j)} - \delta_{ij} \partial_{k} V_{m} (\partial_{k} V_{m}
- \partial_{m} V_{k}) - \frac{3}{4} \delta_{ij} (\partial_{t} V)^{2} \right].
\end{eqnarray}\end{subequations}}\noindent
The spatial traces of the potentials will be denoted by
$\hat{W}\equiv\hat{W}_{ii}$ and $\hat{Z}\equiv\hat{Z}_{ii}$. The harmonic
gauge condition of the near-zone PN expansion results in four independent PN
differential identities to be satisfied by the latter potentials,
\begin{subequations}\label{hgauge}\begin{eqnarray}
&& \partial_{t} \left\{ V + \frac{1}{c^{2}} \left[\frac{1}{2}\hat{W} + 2 V^{2}
\right] + \frac{4}{c^{4}} \left[\hat{X} + \frac{1}{2} \hat{Z} + \frac{1}{2} V
\hat{W} + \frac{2}{3} V^{3} \right] \right\} \nonumber\\ && \qquad +
\partial_{i} \left\{ V_{i} + \frac{2}{c^{2}} \left[\hat{R}_{i} + V V_{i}
\right] + \frac{4}{c^{4}} \left[ \hat{Y}_{i} - \frac{1}{2} \hat{W}_{ij} V_{j}
+ \frac{1}{2} \hat{W} V_{i} + V\hat{R}_{i} + V^{2} V_{i} \right] \right\} =
\mathcal{O} \left(\frac{1}{c^{6}} \right), \nonumber\\ \\&& \partial_{t}
\left\{ V_{i} + \frac{2}{c^{2}} \left[\hat{R}_{i} + V V_{i}\right] \right\} +
\partial_{j} \left\{ \hat{W}_{ij} - \frac{1}{2} \hat{W} \,\delta_{ij} +
\frac{4}{c^{2}} \left[\hat{Z}_{ij} - \frac{1}{2} \hat{Z} \,\delta_{ij} \right]
\right\} = \mathcal{O} \left(\frac{1}{c^{4}} \right).
\end{eqnarray}\end{subequations}
The harmonicity conditions, Eqs.~(\ref{hgauge}), thus provide a verification
of certain computed potentials which are required at the 3.5PN order. We note,
however, that these conditions are unable to provide a direct check to the
potentials with the most challenging form which are necessary for our
purposes.

As suggested by their definitions (\ref{pot1})--(\ref{pot3}), several
recurring structures may be identified in our retarded potentials. The
advantages of decomposing the potentials are substantial; general schemata are
developed to solve integrals of a particular form, which not only simplify
considerably the computational aspect but also reveal interesting analytical
solutions. The potentials comprise essentially three types of hierarchical
terms of increasing complexity \cite{BFP98,BFeom}:

\begin{enumerate}
\item[a)] \textit{Compact} (C) potentials involve spatially compact source
terms, proportional to the mass, current and stress densities, $\sigma$,
$\sigma_{i}$ and $\sigma_{ij}$. The support of the source of these potentials
is limited to the domain of the matter system, which will be given in this
instance (the black hole ``particle-model''), by $\delta$-function
singularities (see Section \ref{secIII}). We have, e.g.,
\begin{subequations}\label{compactpot}\begin{eqnarray}
V^{\mathrm{(C)}} & \equiv & V = \Box_{\mathcal{R}}^{-1}\left[-4 \pi
G\, \sigma\right],\\ \hat{X}^{\mathrm{(C)}} & = &
\Box_{\mathcal{R}}^{-1}\left[-4 \pi G\, V \sigma_{ii}\right],\\
\hat{W}_{ij}^{\mathrm{(C)}} & = & \Box_{{\cal R}}^{-1}\left[-4 \pi G\,
(\sigma_{ij} - \delta_{ij} \sigma_{kk})\right].
\end{eqnarray}\end{subequations}
The C potentials are relatively simple, but, for instance, $V^{\mathrm{(C)}}$
must be calculated with the full 3.5PN precision.

\item[b)] \textit{Quadratic non-compact} (QNC) potentials are generated by
spatially non-compact supported distribution of the (source-induced)
gravitational field. Specifically, such potentials include terms of the
symbolic type $\sim \Box_{\mathcal{R}}^{-1} \partial V \partial V $, which
denote the quadratic product of two compact potentials $V$ and/or $V_{i}$ and
their space-time derivatives. Examples of QNC potential terms are,
\begin{subequations}\label{quadpot}\begin{eqnarray}
\hat{W}_{ij}^{\mathrm{(QNC)}} & = & \Box_{{\cal R}}^{-1}\left[ - \partial_i V
\partial_j V \right],\\ \hat{Z}_{ij}^{\mathrm{(QNC)}} & = & \Box_{\mathcal{R}}^{-1}
\left[ - 2 \partial_{(i} V \partial_{t} V_{j)} + \partial_{i} V_{k}
\partial_{j} V_{k} + \partial_{k} V_{i}\partial_{k} V_{j} \right.\nonumber\\
&& \qquad \left. - 2 \partial_{(i} V_{k} \partial_{k} V_{j)} - \delta_{ij}
\partial_{k} V_{m} (\partial_{k} V_{m} - \partial_{m} V_{k}) - \frac{3}{4}
\delta_{ij} (\partial_{t} V)^{2} \right],\\ \hat{X}^{\mathrm{(QNC)}} & = &
\Box_{\mathcal{R}}^{-1}\left[ \hat{W}_{ij}^{\mathrm{(C)}} \partial^2_{ij} V + 2
V_{i} \partial_{t} \partial_{i} V + V \partial_{t}^{2} V \right.\nonumber\\ &
& \qquad\left. + \frac{3}{2}(\partial_{t} V)^{2} - 2 \partial_{i}
V_{j}\partial_{j} V_{i}\right].
\end{eqnarray}\end{subequations}

\item[c)] \textit{Cubic non-compact} (CNC) potentials include more complicated
integral expressions of the form of the product of a quadratic QNC potential
and a compact C potential. The symbolic form reads $\sim
\Box_{\mathcal{R}}^{-1}\left[\Box_{\mathcal{R}}^{-1}( \partial V \partial V
)\,\partial V\right]$. The paradigm of such terms is,
\begin{equation}\label{cubpot}
\hat{X}^{\mathrm{(CNC)}} =
\Box_{\mathcal{R}}^{-1}\left[\hat{W}_{ij}^{\mathrm{(QNC)}}
\partial^2_{ij} V \right],
\end{equation}
which must be evaluated at relative 1.5PN order. The other CNC terms, present
in the potentials $\hat{T}$ and $\hat{Y}_i$, will only need to be controlled at the 
0.5PN order.
\end{enumerate}

\subsection{The equations of motion in terms of the potentials}\label{secIIb}

The 3.5PN equations of motion in the case of a smooth hydrodynamical fluid are
obtained by replacing the expression of the metric (\ref{metricg}) into the
law of covariant conservation of the matter stress-energy tensor, $\nabla_\nu
T^{\mu\nu}=0$, which is equivalent to the equation of geodesics in the case of
point particle sources. The resulting equation can be expressed as,
\begin{equation}\label{eqofm1}
\frac{dP^{i}}{dt} = F^{i},
\end{equation}
where $P^{i}$ and $F^{i}$, introduced here for convenience, can be thought of as
some effective \textit{linear momentum density} and \textit{force density} of
the matter system respectively, and are defined by,
\begin{subequations}\label{linearmomdensity}\begin{eqnarray}
P^{i} & = & \frac{g_{i\mu} v^{\mu}}{\sqrt{ - g_{\rho\sigma}
\frac{v^{\rho} v^{\sigma}}{c^{2}}}},\\ F^{i} & = & \frac{1}{2}
\frac{\partial_{i} g_{\mu\nu} v^{\mu} v^{\nu}}{ \sqrt{ -
g_{\rho\sigma} \frac{v^{\rho} v^{\sigma}}{c^{2}}}}.
\end{eqnarray}\end{subequations}
Here, $v^i=dx^i/dt$ denotes the \textit{coordinate} velocity field ($t=x^0/c$),
and we pose $v^{\mu}=(c,v^i)$. Substituting the metric (\ref{metricg}) into
the above expressions (\ref{linearmomdensity}), and performing the PN
re-expansion, gives \cite{BFeom},
\allowdisplaybreaks{\begin{subequations}\label{eqnmotion1}\begin{eqnarray} P^i
&=& v^i \nonumber\\ &+& \frac{1}{c^2}\, \left( \frac{1}{2} v^2 v^i + 3 V v^i -
4 V_i \right) \nonumber\\ &+& \frac{1}{c^4}\, \left(\frac{3}{8} v^4 v^i +
\frac{7}{2} V v^2 v^i \ - \ 4 V_j v^i v^j - 2 V_i v^2 \right. \nonumber\\ &&
\qquad + \left. \frac{9}{2} V^2 v^i - 4 V V_i + 4 \hat{W}_{ij} v^j - 8
\hat{R}_i \right) \nonumber\\ &+& \frac{1}{c^6} \left( \frac{5}{16} v^6 v^i +
\frac{33}{8} V v^4 v^i - \frac{3}{2} V_i v^4 - 6 V_j v^i v^j v^2 +
\frac{49}{4} V^2 \, v^2 v^i \right. \nonumber\\ && \qquad + 2 \hat{W}_{ij} v^j
v^2 + 2 \hat{W}_{jk} v^i v^j v^k - 10 V V_i v^2 - 20 V V_j v^i v^j \nonumber\\
&& \qquad - 4 \hat{R}_i v^2 - 8 \hat{R}_j v^i v^j + \frac{9}{2} V^3 v^i + 12
V_j V_j v^i + 12 \hat{W}_{ij} V v^j \nonumber\\ && \qquad + 12\hat{X} v^i + 16
\hat{Z}_{ij} v^j - 10 V^2 V_i \nonumber\\ && \qquad \left. - 8 \hat{W}_{ij}
V_j - \ 8 V \hat{R}_i - 16 \hat{Y}_i \right) \, + \mathcal{O} \left(
\frac{1}{c^8} \right),\\ F^i &=& \partial_i V \nonumber\\ &+& \frac{1}{c^2}
\left( - V \, \partial_i V + \frac{3}{2} \partial_i V \, v^2 - 4 \partial_i
V_j \, v^j \right)\nonumber\\ &+& \frac{1}{c^4}\, \left( \frac{7}{8}
\partial_i V \, v^4 - 2 \partial_i V_j \, v^j v^2 + \frac{9}{2} V \,
\partial_i V \, v^2 + 2 \partial_i \hat{W}_{jk} \, v^j v^k - 4 V_j \,
\partial_i V \, v^j \right. \nonumber\\ && \qquad - 4 V \, \partial_i V_j \,
v^j - \left. 8 \partial_i \hat{R}_j \, v^j + \frac{1}{2} V^2\, \partial_i V \,
+ 8 V_j \, \partial_i V_j + 4 \partial_i \hat{X} \right) \nonumber\\ &+&
\frac{1}{c^6} \left( \frac{11}{16} v^6 \partial_i V - \frac{3}{2} \partial_i
V_j \, v^j v^4 + \frac{49}{8} V \, \partial_i V \, v^4 + \partial_i
\hat{W}_{jk} \, v^2 v^j v^k \right. \nonumber\\ && \left. \qquad - 10 V_j \,
\partial_i V\, v^2 v^j - 10 V \partial_i V_j \, v^2 v^j - 4 \partial_i
\hat{R}_{j}\, v^2 v^j + \frac{27}{4} V^2 \,\partial_i V \, v^2 \right.
\nonumber\\ && \left. \qquad + 12 V_j \partial_i V_j\, v^2 + 6 \hat{W}_{jk} \,
\partial_i V\, v^j v^k + 6 V \, \partial_i \hat{W}_{jk} \, v^j v^k + 6
\partial_i \hat{X} v^2 \right. \nonumber\\ && \left. \qquad + 8 \partial_i
\hat{Z}_{jk}\, v^j v^k - 20 V_j V \partial_i V v^j - 10 V^2 \, \partial_i V_j
\, v^j - 8 V_k \, \partial_i \hat{W}_{jk} \, v^j \right. \nonumber\\ && \left.
\qquad - 8 \hat{W}_{jk} \, \partial_i V_k \, v^j - 8 \hat{R}_{j} \,
\partial_i V \, v^j - 8 V \, \partial_i \hat{R}_{j}\, v^j - 16 \partial_i
\hat{Y}_j\, v^j \right. \nonumber\\ && \left. \qquad - \frac{1}{6} V^3\,
\partial_i V - 4 V_j \, V_j \, \partial_i V \, + 16 \hat{R}_{j} \, \partial_i
V_j \, + 16 V_j \, \partial_i \hat{R}_j \right.\nonumber\\ && \left. \qquad -
8 V\, V_j\, \partial_i V_j - 4 \hat{X}\, \partial_i V - 4 V \, \partial_i
\hat{X} \, + 16 \partial_i \hat{T} \right) \, + \mathcal{O} \left(
\frac{1}{c^8} \right).
\end{eqnarray}\end{subequations}}\noindent
From this, we deduce the coordinate acceleration as,
\begin{equation}\label{eqnmotion2}
a^i = F^i - \frac{d}{dt}\left(P^i-v^i\right).
\end{equation}

The radiation reaction terms in the equations of motion,
(\ref{eqnmotion1})--(\ref{eqnmotion2}), appear explicitly when performing the
PN expansion of our elementary non-linear potentials. They are obtained by the
careful consideration of all possible contributions at the 2.5PN or 3.5PN
orders in each of these potentials. More precisely, the 2.5PN and 3.5PN
contributions arise both from the expansion of the retardation of the inverse
d'Alembertian retarded integrals, and from the PN corrections already present
in the sources of the potentials. For instance, the sources,
Eqs.~(\ref{pot1})--(\ref{pot3}), involve the mass, current and stress
densities, $\sigma$, $\sigma_{i}$ and $\sigma_{ij}$ given by (\ref{sigma}),
which depend themselves on the potentials in a manner consistent with the
iterative PN formalism. In addition, some required contributions also occur from
the systematic \textit{order reduction} of the accelerations, which is applied
when calculating the time derivatives associated with the retardations of the
potentials, or for instance, when performing the total time derivative of the
linear momentum density function, $P^{i}$ in Eq.~(\ref{eqnmotion2}). By order
reduction, we refer to the replacement of the acceleration by its explicit
expression given by the PN equations of motion in terms of the bodies'
positions and velocities, and the subsequent PN re-expansion, which is
performed in a consistent manner at the PN order in question.

From Eqs.~(\ref{eqnmotion1})--(\ref{eqnmotion2}), it is apparent that the
expressions for all the potentials or their spatial derivatives are required
at a relative 0.5PN order above the existing explicit 3PN expansions computed
in \cite{BFeom}. They must be expressed solely as a function of the masses and
velocities (after the order-reduction of the accelerations). For the moment,
Eqs.~(\ref{eqnmotion1})--(\ref{eqnmotion2}) were derived in the case of
general matter systems, and the precise mathematical description of the
source, \textit{i.e.} the matter stress-energy tensor $T^{\mu \nu}$, is
required next. Section \ref{secIII} introduces the $\delta$-function model of
the compact binary and the associated regularisation.

The PN precision in each of the potentials and their gradients, which we
require in order to control the 3.5PN acceleration, is given in the following
Table. It is convenient to distinguish between the computation of a potential
and the one of its gradient, because in the case of the odd terms, the
gradient is often easier to compute. This is due to the simplification that
the 0.5PN relative term in the expansion of the retardation of a potential (we
refer later to such an odd-parity term as ``retardation-like'') is always a
mere function of time, which thus vanishes when taking the gradient. On the
other hand, the gradient of a potential is often required at a higher PN order
than the potential itself, so it is generally good practice to perform a
separate computation for the gradient. At 3.5PN order, it is necessary to
develop the odd terms in,
\begin{center}\label{table}
\begin{tabular}{p{3.0cm}cp{4cm}p{3.0cm}cp{4cm}}
$\partial_i V$ & to order & ~$\mathcal{O}(c^{-7})$, & $\partial_i \hat{X}$ &
to order & ~$\mathcal{O}(c^{-3})$,\\ $V$ & \textit{id.} &
~$\mathcal{O}(c^{-5})$, & $\hat{X}$ & \textit{id.} & ~$\mathcal{O}(c^{-1})$,\\
$V_i$ and $\partial_j V_i$ & \textit{id.} & ~$\mathcal{O}(c^{-5})$, &
$\hat{Z}_{ij}$ and $\partial_k \hat{Z}_{ij}$ & \textit{id.} &
~$\mathcal{O}(c^{-1})$,\\ $\hat{W}_{ij}$ and $\partial_k \hat{W}_{ij}$ &
\textit{id.} & ~$\mathcal{O}(c^{-3})$, & $\hat{Y}_i$ and $\partial_j
\hat{Y}_{i} $ & \textit{id.} & ~${\cal O}(c^{-1})$,\\ $\hat{R}_{i}$ and
$\partial_j \hat{R}_{i}$ & \textit{id.} & ~$\mathcal{O}(c^{-3})$, &
$\partial_i \hat{T}$ & \textit{id.} & ~${\cal O}(c^{-1})$.\\
\end{tabular}
\end{center}
Apart from the purely compact support potentials, $V$ and $V_{i}$, the above
potentials consist of both compact (C) and non-compact (QNC and/or CNC)
support distributed sources. The following Sections \ref{secIII}--\ref{secIV}
systematically treat how to evaluate each of these different types of
contributions.
We shall find that the evaluation of the required integrals for the QNC and
CNC terms yield explicit \textit{closed-form} expressions, valid at any
field point over all space, for all the odd parts of potentials in the
previous Table. This is in contrast with the computation of the equations of
motion at the previous 3PN order \cite{BFeom}, where closed-form solutions to
certain non-linear Poisson-like integrals could not be given at any field
point but existed only at the location of each particle (in a regularised
sense). As Section \ref{secIV} illustrates, the alternative ``direct'' evaluation
method of Poisson-like integrals at the location of the particles, using the
same method as for the 3PN equations of motion \cite{BFreg,BFeom}, provides a
further verification of our analytic closed-form Poisson-like solutions.

\section{Application to point particles}\label{secIII}

The compact binary system is modelled as two structureless point particles
with masses $m_{1}$ and $m_{2}$, which are described by $\delta$-function
singularities, and move on a general, not necessarily circular, orbit. We
neglect the intrinsic rotations (spins) of the particles.\,\footnote{If
necessary, the spins may be added to the formalism along the lines of
Refs.~\cite{BOC79,KWWi93,TOO01}.} As part of the formalism of \cite{BFeom}, we
assume that although the equations of motion
(\ref{eqofm1})--(\ref{eqnmotion2}) were derived under the assumption of a
general smooth stress-energy tensor, \textit{i.e.}
$C^{\infty}(\mathbb{R}^{3})$, they remain valid in the case of point
particles, provided that we supplement the calculation by a consistent use of
a self-field regularisation. As discussed in the Introduction, the Hadamard
self-field regularisation is appropriate for the present purpose.

The use of the point particle model is physically justified in the case of
\textit{compact} objects by the fact that, using a Newtonian argument, the
tidal effects are formally equivalent to a correction of the order 5PN
$\sim\mathcal{O}(c^{-10})$ compared to the Newtonian force law (see
\textit{e.g.} \cite{Bliving}). An \textit{a posteriori} justification is also
that the 2.5PN equations of motion of self-gravitating, extended compact
bodies, as derived in Refs.~\cite{GKop86,Kop85,IFA01,PW02}, are in complete
agreement with those derived for the model of point-particles in
\cite{DD81a,D82,D83houches,BFP98}. The same is true at 3PN order: there is
agreement between the 3PN equations of motion of extended compact bodies
\cite{itoh1,itoh2} and those for point particles \cite{BFeom,BDE04}. A general
way to justify the use of structureless point-particles in order to describe
gravitationally condensed objects is to invoke the ``effacing property'' of
General Relativity (a consequence of the strong version of the equivalence
principle). According to this property, the internal structure is effaced when
considering the motion and the radiation of the compact bodies, so that one
can describe them only by their masses \cite{D83houches}. Once the use of
delta functions is physically justified, the advantage, of course, is that
they considerably simplify the calculations.

Following the standard prescription in General Relativity, we write the
distributional stress-energy tensor of point particles as,
\begin{equation}\label{model1}
T^{\mu\nu}(\mathbf{x},t) = \mu_{1} (t) \,v_{1}^{\mu} (t) \,v_{1}^{\nu} (t)
\,\delta ( \mathbf{x} - \mathbf{y}_1 (t)) + 1 \,\leftrightharpoons\, 2,
\end{equation}
where $\mathbf{x}$ is the field point, $\mathbf{y}_1$ and $\mathbf{y}_2$ are
the positions of the particles, $\mathbf{v}_1(t) = d\mathbf{y}_1(t)/dt$ is the
coordinate velocity in harmonic coordinates, and $v_{1}^{\mu} \equiv
(c,\mathbf{v}_1)$. The symbol $1 \,\leftrightharpoons\, 2$ means the same
terms but with the particles' labels 1 and 2 exchanged, and $\delta$ denotes
the usual Dirac three-dimensional delta function. Alternatively, for the mass,
current and stress densities defined by Eqs.~(\ref{sigma}), we find,
\begin{subequations}\label{densitesigmapart}\begin{eqnarray}
\sigma & = & \widetilde{\mu}_{1} \,\delta(\mathbf{x}-\mathbf{y}_1) + 1
\,\leftrightharpoons\, 2, \label{sigmapart}\\ \sigma_{i} & = & \mu_{1}
\,v_{1}^{i}\,\delta(\mathbf{x}-\mathbf{y}_1) + 1 \,\leftrightharpoons\, 2, \\
\sigma_{ij} & = & \mu_{1}
\,v_{1}^{i}v_{1}^{j}\,\delta(\mathbf{x}-\mathbf{y}_1) + 1
\,\leftrightharpoons\, 2,
\end{eqnarray}\end{subequations}
where the quantities $\mu_{1}$ and $\widetilde{\mu}_{1}$ are some explicit
functions of coordinate time $t$, through the source trajectories,
$\mathbf{y}_{1}(t)$ and $\mathbf{y}_{2}(t)$, and velocities,
$\mathbf{v}_{1}(t)$ and $\mathbf{v}_{2}(t)$, given by,
\begin{subequations}\label{mumutilde}\begin{eqnarray}
\label{mu}\mu_{1}(t) &=& \frac{m_{1}}{\sqrt{(g\,g_{\rho\sigma})_{1}
\frac{v_{1}^{\rho}v_{1}^{\sigma}}{c^{2}}}},\\
\label{mutilde}
\widetilde{\mu}_{1}(t) &=& \mu_{1}(t)\left[1 +
\frac{\mathbf{v}_1^{2}}{c^{2}}\right].
\end{eqnarray}\end{subequations}
Here, $(g_{\rho\sigma})_{1}$ and $(g)_{1}$ denote the values of the metric and
its determinant computed at the position of the particle 1 following the
prescription of the Hadamard partie-finie introduced in Eq.~(\ref{F1}). As we
emphasise later, we do not encounter any problems associated with the
``non-distributivity'' of Hadamard's regularisation at the present 3.5PN
order; so, for instance, $(g\,g_{\rho\sigma})_{1}$ in Eq.~(\ref{mumutilde})
may be replaced by the product of regularisations,
$(g)_{1}\,(g_{\rho\sigma})_{1}$. Thus, it is unnecessary
to take into account the subtleties associated with several possible choices
for the stress-energy tensor in the context of Hadamard's regularisation,
which depend on whether the factors of the delta-function
$\delta(\mathbf{x}-\mathbf{y}_1)$ are supposed to be evaluated at any field
point $\mathbf{x}$ or at the particle's position $\mathbf{y}_1$ as assumed in
Eqs.~(\ref{mumutilde}).\,\footnote{For instance, a different prescription,
valid in the context of the extended Hadamard regularisation, was given in
Ref.~\cite{BFregM}. In this prescription, the factor in Eq.~(\ref{mu})
involving the determinant of the metric is calculated at $\mathbf{x}$ whilst
the other factor is evaluated at $\mathbf{y}_1$; in addition, a special
version of the Dirac delta-function, designed in such a way that it permits to
keep track of the Lorentz invariance of the formalism, is assumed in
\cite{BFregM}.} The latter problems resulted in the appearance of some
ambiguities in the application of Hadamard's regularisation at the 3PN order.
The ambiguities have since then been resolved by means of dimensional
regularisation \cite{DJSdim,BDE04,BDEI04}, but do not concern us here as we
are interested in the 3.5PN approximation which is unambiguous. We are thus
following here the straightforward prescription for the stress-energy tensor
of point particles in General Relativity.

We must now be more specific on the method chosen for computing the metric
coefficients at the point 1 in Eqs.~(\ref{mumutilde}). Let $F(\mathbf{x})$ be
a typical function we encounter in the problem, where for convenience, we
indicate only the relevant dependence of the function on the field point
$\mathbf{x}$ (it depends also on coordinate time $t$ through the source
positions and velocities). The function $F(\mathbf{x})$ is smooth on
$\mathbb{R}^3$, except at the singular points $\mathbf{y}_{1}$ and
$\mathbf{y}_{2}$, around which it admits a power-like singular expansions of
the type (for any $N \in \mathbb{N}$),
\begin{equation}\label{Fexp}
F(\mathbf{x}) = \sum_{a_0 \le a \le N} \, r^a_1 \,f_1^{[a]}(\mathbf{n}_1) +
\mathcal{O}\left(r^N_1\right),
\end{equation}
(and similarly when $1\,\leftrightharpoons\, 2$), where
$r_{1}\equiv\vert\mathbf{x}-\mathbf{y}_{1}\vert\rightarrow 0$ and the
coefficients $f_1^{[a]}$ of the various powers of $r_{1}$ depend on the unit
direction of approach to the singularity, $\mathbf{n}_{1}\equiv
(\mathbf{x}-\mathbf{y}_1)/r_{1}$. The powers of $r_{1}$ are relative integers,
$a\in\mathbb{Z}$, and bounded from below by some typically negative integer
$a_{0}$, depending on the $F$ in question. The coefficients $f_1^{[a]}$ for
which $a< 0$ are called the \textit{singular} coefficients of $F$. The class
of functions such as $F$ is called $\mathcal{F}$. The Hadamard \textit{partie
finie} of $F\in\mathcal{F}$ at the singular point 1 is then defined by the
angular average,
\begin{equation}\label{F1}
(F)_1 = \int \frac{d \Omega_1}{4 \pi} f_1^{[0]}(\mathbf{n}_1),
\end{equation}
where $d \Omega_1\equiv d\Omega (\mathbf{n}_{1})$ is the solid angle which is
centered on $\mathbf{y}_{1}$ and in the direction $\mathbf{n}_{1}$. In
principle, the Hadamard partie finie is non-distributive (with respect to
multiplication) in the sense that $(FG)_{1}\neq (F)_{1}(G)_{1}$ in general for
$F$ and $G$ belonging to $\mathcal{F}$. However, for the terms occurring at
3.5PN order, it is unnecessary to account for this feature, as all the
functions encountered at this order will in fact be such that
$(FG)_{1}=(F)_{1}(G)_{1}$.

The Hadamard partie finie of an integral, in short $\mathrm{Pf}\int d^3
\mathbf{x}\,F$, which has divergencies due to the singular expansion
(\ref{Fexp}) of the function around at the singular points $\mathbf{y}_{1}$
and $\mathbf{y}_{2}$, is defined by the always existing limit,
\begin{eqnarray}\label{PfF}
\mathrm{Pf} \int d^3 \mathbf{x}\,F & = & \lim_{s \to 0} \left\{
\int_{\mathcal{S}(s)} d^3 \mathbf{x}\,F(\mathbf{x}) \right. \nonumber\\ &&
\qquad\left. + \sum_{a+3<0} \frac{s^{a+3}}{a+3} \int d
\Omega_1\,f_1^{[a]}(\mathbf{n}_1) + \ln \left( \frac{s}{s_1} \right) \int d
\Omega_1\,f_1^{[-3]}(\mathbf{n}_1) \right. \nonumber\\ && \qquad \left. + 1
\,\leftrightharpoons\, 2 \right\},
\end{eqnarray}
where in the R.H.S., the integration extends on the domain
$\mathcal{S}(s)\equiv\mathbb{R}^3\setminus\mathcal{B}_1(s)\cup\mathcal{B}_2(s)$,
\textit{i.e.} defined by the whole space from which one has excised two
coordinate balls $\mathcal{B}_1(s)$ and $\mathcal{B}_2(s)$ centred on the two
particles and having the (same) radius $s$. The other terms are defined with
the help of the singular coefficients in the expansion of the function given
by (\ref{Fexp}). The regularisation (\ref{PfF}) depends in principle on two
constants, $s_1$ and $s_2$, appearing in the logarithmic terms. These
constants play an important role at 3PN order, but will never appear in the
present work, since there are no logarithmic divergencies at the 3.5PN order.
In Eq.~(\ref{PfF}), we suppose that the integral converges at infinity, when
$\vert\mathbf{x}\vert\rightarrow +\infty$. We shall explain below how one
treats the integral in the case where it diverges at infinity.

In the extended version of Hadamard's regularisation \cite{BFreg,BFregM}, one
associates to any $F \in \mathcal{F}$ a partie finie ``pseudo-function'',
\textit{i.e.} a linear form defined on the set $\mathcal{F}$, which permits us
to give a precise meaning to the notions of Dirac delta functions, and
derivatives of singular functions in a distributional sense, when they are
mutiplied by or act on other singular functions in the class $\mathcal{F}$. The
detailed construction of Ref.~\cite{BFreg} was useful at 3PN order, but is not
needed in the present work. Nevertheless, it is convenient, because of the
availability of the computer programmes used in \cite{BFeom}, to adopt all the
rules of the extended Hadamard regularisation (we know anyway that the
different variants of Hadamard's regularisation give the same answer at 3.5PN
order). In particular, the Dirac delta-function is defined in the extended
Hadamard regularisation by (for any $F\in\mathcal{F}$),
\begin{equation}
\label{Pfdelta1}
\mathrm{Pf} \int d^3 \mathbf{x}\,\delta (\mathbf{x}-\mathbf{y}_1)F(\mathbf{x})
= (F)_1,
\end{equation}
where $(F)_1$ is the partie finie of the function given by (\ref{F1}), and
where the indication $\mathrm{Pf}$ reminds us that the equality is true in the
sense of the partie finie integral (\ref{PfF}).\,\footnote{The ``partie finie
delta-function'', satisfying (\ref{Pfdelta1}), has been developed in
Ref.~\cite{BFreg} by the limiting case of a particular class of
pseudo-functions defined from the notion of the Riesz delta function
\cite{Riesz}.}

In addition, the derivatives of singular functions, say $\partial_iF$, are to
be performed in a distributional sense, and for the present work, we use the
explicit formula of the extended Hadamard regularisation,
\begin{equation}
\label{distderiv}
\partial_i F = \left(\partial_i F\right)_\mathrm{ordinary} + \mathrm{D}_i[F],
\end{equation}
where the first term represents the derivative in the ordinary sense (as
algebraic computer programs would compute it), and where the second term is
the purely distributional part of the derivative, given explicitly by,
\begin{equation}
\label{distderiv1}
\mathrm{D}_i[F] = 4 \pi\, n^i_1 \left[ \frac{1}{2}
\,r_1\,f_1^{[-1]}(\mathbf{n}_1) + \sum_{k \geq 0}
\frac{1}{r^k_1}\,f_1^{[-2-k]}(\mathbf{n}_1) \right]
\delta(\mathbf{x}-\mathbf{y}_1) + 1 \,\leftrightharpoons\, 2.
\end{equation}
The distributional terms depend only on the singular coefficients of the
expansion of $F$. It was shown in Ref.~\cite{BFreg} that this derivative
generalises the usual distributional derivative of distribution theory in the
context of the class of singular functions $\mathcal{F}$. It is such that one
can integrate by parts any integrals; in particular, the integral of the
gradient of any $F\in\mathcal{F}$, considered in the previous distributional
sense, is always zero. Multiple derivatives, as well as time derivatives, are
treated in a similar way, and the reader is referred to \cite{BFreg} for
details. Notice, however, that the distributional derivative
(\ref{distderiv})--(\ref{distderiv1}), like the usual distributional
derivative of distribution theory \cite{Schwartz}, is seen not to satisfy the
Leibniz rule for the derivation of a product. This poses a problem at the 3PN
order (by the presence of certain ambiguity parameters, later resolved by
means of dimensional regularisation), but not at the next order of 3.5PN.

In our investigations, we shall always consider singular functions
$F\in\mathcal{F}$ in the form of a PN expansion. In order to be systematic, we
introduce a special notation for the PN coefficients (with odd or even-type
parity) of the function $F$, say,
\begin{equation}\label{PNF}
F(\mathbf{x},t,c)=\sum_n\frac{1}{c^n}\mathop{F}_{(n)}(\mathbf{x},t).
\end{equation}
In the present paper, we neglect the dependence of the PN coefficients on the
logarithm of $c$, since such $\ln c$ terms do not occur at the 3.5PN order. The
coefficients $\mathop{F}_{(n)}$ (and their gradients) will be computed at the
points 1 or 2 by means of the partie finie (\ref{F1}), leading to the
evaluation of such objects like,
\begin{equation}\label{PNF1}
(\mathop{F}_{(n)})_1 = \int \frac{d \Omega_1}{4 \pi}
\mathop{f}_{(n)}{}_{\!\!1}^{\!\![0]}(\mathbf{n}_1),
\end{equation}
where $F$ stands for any of the PN iterated potentials $V$, $V_i$,
$\hat{W}_{ij}$, $\cdots$ defined in Section
\ref{secII}.

\section{Computation of the non-linear potentials}\label{secIV}

\subsection{Compact-support potentials}\label{secIVa}

Following the nomenclature convention introduced in Section \ref{secII},
compact (C) support potentials refer to inverse d'Alembertian retarded
integrals of some source terms, which possess as a factor, the source mass,
current or stress densities defined by Eqs.~(\ref{sigma}); the latter,
therefore, for our model of point particles, are of the form $F_{1}\,\delta_1$ and
$1\,\leftrightharpoons\, 2$, where $F_{1}(\mathbf{x})\in\mathcal{F}$ and
$\delta_1(\mathbf{x})\equiv\delta (\mathbf{x}-\mathbf{y}_1)$. We shall
illustrate the scheme by the derivation of the 3.5PN term in the purely
compact-support potential $V\equiv V^\mathrm{(C)}$ defined by Eq.~(\ref{V}).
Performing the expansion of retardations inside the d'Alembertian integral, we
obtain,
\begin{equation}\label{exppotV}
V(\mathbf{x}',t) = G
\sum_{n=0}^{7}\frac{(-)^n}{n!\,c^{n}}\left(\frac{\partial}{\partial
t}\right)^{n}\int
d^{3}\mathbf{x}\,\vert\mathbf{x}'-\mathbf{x}\vert^{n-1}\sigma(\mathbf{x},t)+
\mathcal{O}\left(\frac{1}{c^{8}}\right).
\end{equation}
The source density, $\sigma$, is substituted by its expression valid for two
point-masses, Eq.~(\ref{sigmapart}), where we recall that
$\widetilde{\mu}_1(t)$ is a function of time defined by (\ref{mumutilde}).
This results in,
\begin{eqnarray}\label{exppotV2}
V & = & G \left\{\frac{\widetilde{\mu}_{1}}{r_{1}} - \frac{1}{c}\,\frac{d}{d
t}\left(\widetilde{\mu}_{1}\right)
+\frac{1}{2\,c^{2}}\,\frac{\partial^2}{\partial
t^2}\left(\widetilde{\mu}_{1}\,r_{1}\right) -
\frac{1}{6\,c^3}\,\frac{\partial^3}{\partial
t^3}\left(\widetilde{\mu}_{1}\,r_{1}^{2}\right) +
\frac{1}{24\,c^{4}}\,\frac{\partial^4}{\partial
t^4}\left(\widetilde{\mu}_{1}\,r_{1}^{3}\right) \right.\nonumber\\ && \qquad
\left. - \frac{1}{120\,c^{5}}\,\frac{\partial^5}{\partial
t^5}\left(\widetilde{\mu}_{1}\,r_{1}^{4}\right) +
\frac{1}{720\,c^{6}}\,\frac{\partial^6}{\partial
t^6}\left(\widetilde{\mu}_{1}\,r_{1}^{5}\right) -
\frac{1}{5040\,c^{7}}\,\frac{\partial^7}{\partial
t^7}\left(\widetilde{\mu}_{1}\,r_{1}^{6}\right) \right\}\nonumber\\ &&  + 1
\,\leftrightharpoons\, 2 +
\mathcal{O}\left(\frac{1}{c^{8}}\right),
\end{eqnarray}
where $r_1=\vert\mathbf{x}-\mathbf{y}_1\vert$ (for convenience we call
$\mathbf{x}$ the field point in this formula). Notice that the $1/c$ term is a
mere function of time, and thus vanishes when taking the spatial gradient
(hence we employ for this term the notation for a total time derivative $d/dt$
instead of the partial derivative $\partial/\partial t$). Furthermore, one can
see that this term is actually of order $1/c^3$ since the Newtonian
approximation to $\widetilde{\mu}_{1}$, namely $m_1$, is constant. Performing
repeatedly the time derivatives of $r_1$ introduces some accelerations which
must be consistently order reduced by means of the PN equations of motion. In
Eq.~(\ref{exppotV2}), we require the expression of $\widetilde{\mu}_1(t)$ up
to 3.5PN order, which we easily find by inserting the PN metric
(\ref{metricg}) into the definition (\ref{mumutilde}), and results in,
\begin{eqnarray}\label{mutildeexplicit}
\frac{\widetilde{\mu}_1}{m_1} &=& 1 \nonumber\\ &+& \frac{1}{c^2} \left[ -
(V)_1 + \frac{3}{2} \,v^2_1\right] \nonumber\\ &+& \frac{1}{c^4} \left[ - 2
\,(\hat{W})_1 + \frac{1}{2}\,(V^2)_1 + \frac{1}{2}\,(V)_1\,v_1^2 - 4 \,(V_i)_1
\,v^i_1 + \frac{7}{8} \,v^4_1 \right]\nonumber\\ &+& \frac{1}{c^6} \left[- 8
\,(\hat{Z})_1 - 4 \,(\hat{X})_1 + 2 \,(\hat{W})_1\,(V)_1 - 4
\,(V_{i})_1\,(V_{i})_1 - \frac{1}{6}\,(V^3)_1 \right.\nonumber\\ &&\quad
\left. + \frac{11}{4}\,(V^2)_1\,v_1^2 - 8 \,(\hat{R}_i)_1 \,v_1^i + 2
(\hat{W}_{ij})_1 v_1^{i} v_1^{j} - 3 \,(\hat{W})_1 \,v_1^{2}
\right.\nonumber\\ &&\quad \left. - 4 \,(V)_1 \,(V_{i})_1 \,v_1^i - 10
\,(V_i)_1 \,v_1^i v_1^2 + \frac{33}{8} (V)_1 v_1^4 + \frac{11}{16} \,v_1^6
\right] \nonumber\\ &+& \mathcal{O}\left( \frac{1}{c^8} \right).
\end{eqnarray}
Here, the value of each of the elementary potentials is taken at the
singularity 1 following the Hadamard partie finie (\ref{F1}). For all the
terms in Eq.~(\ref{mutildeexplicit}), the distributivity of Hadamard's partie
finie at this order is verified, \textit{e.g.}
$(\hat{W}\,V)_1=(\hat{W})_1\,(V)_1$.

Evidently, the computation of $\widetilde{\mu}_1$ and $V$ proceeds using the
PN iteration, where one begins with $V$ at Newtonian order, given by [using
the notation (\ref{PNF})--(\ref{PNF1})],
\begin{subequations}\begin{eqnarray}
\mathop{V}_{(0)} &=& \frac{G\,m_1}{r_{1}}+ 1 \,\leftrightharpoons\, 2,\\
(\mathop{V}_{(0)})_1 &=& \frac{G\,m_2}{r_{12}},
\end{eqnarray}\end{subequations}
where $r_{12}\equiv\vert\mathbf{y}_1-\mathbf{y}_2\vert$. This is then inserted
into the 1PN term of (\ref{mutildeexplicit}) in order to obtain
$\widetilde{\mu}_1$ at 1PN order, which hence enables one to deduce $V$ itself
at 1PN order, and so on. By taking into account the fact that there is no odd
term in $\widetilde{\mu}_1$ at order $1/c^3$, we find that the first odd term
in $V$ at the level $1/c^3$ is given by,
\begin{equation}\label{V3}
\mathop{V}_{(3)} = G \biggl\{ - \frac{d}{d
t}(\mathop{\widetilde{\mu}}_{(2)}{}_{\!\!1})- \frac{1}{6}
\frac{\partial^3}{\partial
t^3}(\mathop{\widetilde{\mu}}_{(0)}{}_{\!\!1}\,r_1^2)\biggr\}+ 1
\,\leftrightharpoons\, 2,
\end{equation}
(where $\mathop{\widetilde{\mu}}_{(0)}{}_{\!\!1}\,=\,m_{1}$). However, the
dominant odd term in the gradient $\partial_iV$ is only at order $1/c^5$.
Notice that there is an odd term $\sim 1/c$ in the case of $\hat{W}_{ij}$ and
for all the potentials besides $V$ and $V_i$. The computation of both
$(\hat{W})_1$ at 0.5PN order and $(V)_1$ at 1.5PN order are required to get
$\widetilde{\mu}_1$ at 2.5PN order, and we have,
\begin{equation} 
\mathop{\widetilde{\mu}}_{(5)}{}_{\!\!1} = m_1 \biggl\{ - (\mathop{V}_{(3)})_1
- 2 (\mathop{\hat{W}}_{(1)})_1\biggr\}.
\end{equation} 

A useful feature of the potential $\hat{W}\equiv\hat{W}_{ii}$ is that it can
be expressed at the 0.5PN order in a simple way using the following
compact-support form [easily deduced from (\ref{Wij})],
\begin{equation}\label{wii}
\hat{W} = \Box_{\mathcal{R}}^{-1} \left[8 \pi G \left( \sigma_{ii} -
\frac{1}{2} \sigma V \right) \right] - \frac{1}{2} V^2 +
\mathcal{O}\left( \frac{1}{c^2} \right),
\end{equation}
hence we obtain,
\begin{eqnarray}\label{Wii0.5PN}
\mathop{\hat{W}}_{(1)} &=& 2\,G \frac{d}{dt} \int d^3 \mathbf{x} \left(
\mathop{\sigma}_{(0)}{}_{\!\!ii} - \frac{1}{2} \mathop{\sigma}_{(0)}
\mathop{V}_{(0)} \right)\nonumber\\ &=& 2\,G\,m_1 \frac{d}{dt} \left(v_1^2 -
\frac{1}{2} (\mathop{V}_{(0)})_1 \right)+ 1 \,\leftrightharpoons\, 2.
\end{eqnarray}
The 0.5PN term is a purely spatial integral with compact support; it is only a
function of time, hence $(\mathop{\hat{W}}_{(1)})_1=\mathop{\hat{W}}_{(1)}$.
However, for some more involved potentials, the 0.5PN term will also depend on
space, because of the contribution from the Poisson integral of a
corresponding 0.5PN term in the source of the potential.

Continuing in this manner, and using the explicit computations of the NC
potentials as explained in the following Sections, we then obtain
$\widetilde{\mu}_1$ at the required 3.5PN order,
\begin{eqnarray} 
\mathop{\widetilde{\mu}}_{(7)}{}_{\!\!1} &=& m_1 \left\{ -
(\mathop{V}_{(5)})_1 - 2 (\mathop{\hat{W}}_{(3)})_1 +
(\mathop{V}_{(0)})_1(\mathop{V}_{(3)})_1+\frac{1}{2}
(\mathop{V}_{(3)})_1v_1^2\right.\nonumber\\&&\qquad\left.
-4(\mathop{V}_{(3)}{}_{\!\!i})_1v_1^i -8(\mathop{\hat{Z}}_{(1)})_1
-4(\mathop{\hat{X}}_{(1)})_1
+2(\mathop{V}_{(0)})_1(\mathop{\hat{W}}_{(1)})_1\right.\nonumber\\&&\qquad
\left.-8(\mathop{\hat{ R}}_{(1)}{}_{\!\!i})_1v_1^i+2(\mathop{
\hat{W}}_{(1)}{}_{\!\!ij})_1v_1^iv_1^j-3(\mathop{\hat{
W}}_{(1)})_1v_1^2\right\},
\end{eqnarray} 
which allows the compution of $V$ at this order in a straightforward way,
\begin{eqnarray}\label{V7}
\mathop{V}_{(7)} &=& G
\left\{\frac{1}{r_1}\mathop{\widetilde{\mu}}_{(7)}{}_{\!\!1} - \frac{d}{d
t}(\mathop{\widetilde{\mu}}_{(6)}{}_{\!\!1}) + \frac{1}{2}
\frac{\partial^2}{\partial t^2}(\mathop{\widetilde{\mu}}_{(5)}{}_{\!\!1}\,r_1)
- \frac{1}{6} \frac{\partial^3}{\partial
t^3}(\mathop{\widetilde{\mu}}_{(4)}{}_{\!\!1}\,r_1^2)\right.\nonumber\\&&\left.\qquad
- \frac{1}{120} \frac{\partial^5}{\partial
t^5}(\mathop{\widetilde{\mu}}_{(2)}{}_{\!\!1}\,r_1^4)- \frac{1}{5040}
\frac{\partial^7}{\partial
t^7}(\mathop{\widetilde{\mu}}_{(0)}{}_{\!\!1}\,r_1^6)\right\}+ 1
\,\leftrightharpoons\, 2.
\end{eqnarray}
In fact, we only require the gradient of V at 3.5PN order, and so it is
unnecessary to compute the second term in the R.H.S. of (\ref{V7}), which
vanishes when taking the gradient. All these calculations are systematically
performed using algebraic computer programmes.

\subsection{Quadratic non-compact support potentials}\label{secIVb}

The structure of non-compact (NC) support potentials is that of a
d'Alembertian retarded integral whose ``source'' term has a spatially
non-compact support distribution. The integral is perfectly well defined
provided that some sensitive boundary conditions are given for the decay of
the field at past null infinity (the no incoming radiation condition). However,
when we expand the retardations of the d'Alembertian integral, some
Poisson-like integrals will appear at high PN order, which typically become
divergent due to the boundary of the integral at (spatial) infinity. This is
the well-known problem of divergencies of the PN expansion, which is related
to the near-zone limitation of the validity of the PN expansion.

A solution of the latter problem has recently been proposed in
Ref.~\cite{PB02}. Essentially, the work \cite{PB02} showed that the PN
expansion can in fact be iterated \textit{ad infinitum} by using a particular
solution of the Poisson equation at each step, which constitutes an
appropriate generalisation of the usual Poisson integral with a non-compact
support source. In this specific approach, the source term of the Poisson
integral is multiplied by a factor $\vert\mathbf{x}\vert^B$, where $B$ is a
complex parameter, and the solution is defined by the ``finite part'' in the
Laurent expansion of the Poisson integral when the parameter $B$ tends to
zero. In a more recent work \cite{BFN04}, we have written the end result of
\cite{PB02} in an alternative form, which shows that one proceeds with the
expansion of retardations in the PN algorithm by inserting the factor
$\vert\mathbf{x}\vert^B$ inside the integrand and taking the finite part in
the above sense. This procedure will give the correct result for the
radiation-reaction odd terms up to the 3.5PN order (we are of course still
within our specific approach). Starting from the 4PN level, this procedure
will also have to take into account the appearance of tail contributions in
the radiation reaction (see Refs.~\cite{PB02,BFN04} for the details).

Hence, we compute the PN expansion of any
elementary potential by using such a finite part prescription (denoted
$\mathop{\mathrm{FP}}_{B=0}$ in the following) to cure the problem of
divergencies of the integrals at the boundary at infinity. Let us consider as
an example the computation of the non-compact support part of the potential
$\hat{W}_{ij}$, say,
\begin{equation}\label{Wijdef}
\hat{W}_{ij}^\mathrm{(QNC)} \equiv \Box_{\mathcal{R}}^{-1}[-
\partial_{i} V \partial_{j} V],
\end{equation}
which is required, as we have already seen, up to the relative 1.5PN level. By
expanding the retardations up to this level, we obtain,
\begin{eqnarray}\label{wij}
\hat{W}_{ij}^\mathrm{(QNC)}(\mathbf{x}',t) & = & \mathop{\mathrm{FP}}_{B = 0}
\left\{ \Delta^{-1} \, \left[ - r^B \partial_{i} V \partial_{j} V \right] +
\frac{1}{c^2}\frac{\partial^2}{\partial t^2}\Delta^{-2} \,\left[ - r^B
\partial_{i} V \partial_{j} V \right] \right.\nonumber\\ && - \left.\frac{1}{4
\pi c}\,\frac{d}{d t} \int d^{3}\mathbf{x}
\,\vert\mathbf{x}\vert^{B}\,\partial_{i} V \partial_{j} V - \frac{1}{24 \pi
c^{3}} \frac{\partial^3}{\partial t^3} \int
d^{3}\mathbf{x}\,\vert\mathbf{x}\vert^{B}\,\vert\mathbf{x}'-\mathbf{x}\vert^{2}
\, \partial_{i} V \partial_{j} V \right\} \nonumber\\&& +
\mathcal{O}\left(\frac{1}{c^{4}}\right).
\end{eqnarray}
Once again, this particular way of performing the PN expansion, using the
regularisation $\mathop{\mathrm{FP}}_{B=0}$ ``at infinity'', is justified by
our previous works \cite{PB02,BFN04}. From this expression, one sees that the
odd terms at the level $1/c$ and $1/c^{3}$ will come either from the Poisson
or Poisson-like integrals (which are always even) applied to the odd terms
already present in the corresponding source, or from the odd terms coming
directly from the expansion of the retardations (as applied to the even part
of the source). We shall henceforth refer to the first type of odd-parity
terms as \textit{Poisson-like}, and to the second type as
\textit{retardation-like}; ``Poisson-like'' terms correspond to an even-parity
operator applied to an odd source,\,\footnote{We call it the operator of the
instantaneous potentials in Refs.~\cite{PB02,BFN04}, denoted by
$$\mathcal{I}^{-1}=\sum_{k=0}^{+\infty}\left(\frac{\partial}{c\,\partial
t}\right)^{2k}\Delta^{-1-k}.$$} whilst ``retardation-like'' terms consist of
an odd integral (containing explicitly an odd power of $1/c$ in front) with an
even integrand.

To begin with, we see that up to 1.5PN order in Eq.~(\ref{wij}), the odd terms
are only retardation-like, \textit{i.e.} given by the two terms with explicit
powers of $1/c$ and $1/c^3$ (indeed the first odd term in the gradient
$\partial_i V$ occurs only at 2.5PN order). We now compute these
retardation-like terms; later, at the occasion of more complicated potentials,
we shall also see how to compute the Poisson-like terms. We insert into
Eq.~(\ref{wij}) the precise form for the non-compact support source,
$\partial_i V \partial_j V$, using the PN expansion (\ref{exppotV2}) up to
order 1.5PN,
\begin{eqnarray}
\label{wijexp2}
\nonumber \partial_i V \partial_j V & = & G^2 \widetilde{\mu}_{1}^2
\,\partial_{i} \left(\frac{1}{r_1}\right) \partial_j
\left(\frac{1}{r_1}\right) + G^2 \widetilde{\mu}_{1} \widetilde{\mu}_{2}
\,\partial_i
\left(\frac{1}{r_{1}}\right)\partial_j\left(\frac{1}{r_{2}}\right)\\ \nonumber
&-& \frac{G^2 m_1^2}{c^2} \, \left[ a_1^k \partial_{(i}
\left(\frac{1}{r_{1}}\right) \partial_{j)k} \left(r_1\right) - v_1^k v_1^l
\partial_{(i} \left(\frac{1}{r_1}\right) \partial_{j)kl} \left(r_1\right)
\right]\\ \nonumber &-& \frac{G^2 m_1 m_2}{c^2} \, \left[ a_1^k \partial_{(i}
\left(\frac{1}{r_2}\right) \partial_{j)k} \left(r_1\right) - v_1^k v_1^l
\partial_{(i} \left(\frac{1}{r_{2}}\right) \partial_{j)kl} \left(r_1\right)
\right] + 1\,\leftrightharpoons\, 2 \\ &+&
\mathcal{O}\left(\frac{1}{c^{4}}\right).
\end{eqnarray}
Since $\widetilde{\mu}_1$ is a mere function of time, it will not affect the
subsequent reasonning which deals with the spatial integrations. Simply, in
the end, we replace $\widetilde{\mu}_1$ by its explicit expression at 1.5PN
order as deduced from Eq.~(\ref{mutildeexplicit}). Notice again that
(\ref{wijexp2}) is purely ``even'' up to 2PN order. We have replaced, where
appropriate, the masses $\widetilde{\mu}_1$ and $\widetilde{\mu}_2$ by their
Newtonian values $m_1$ and $m_2$.

We can distinguish two types of terms in (\ref{wijexp2}); as suggested by
their names, ``self'' terms refer solely to a single particle and are
proportional to \textit{e.g.} $\widetilde{\mu}_1^2$ or $m_1^2$, whereas
``interaction'' terms are functions of both particles and involve for instance
the product $\widetilde{\mu}_1 \widetilde{\mu}_2$. Furthermore, once the
interaction terms are known, the ``self'' terms can easily be deduced by
taking the limit $\mathbf{y}_{2}\rightarrow\mathbf{y}_{1}$ (and $1
\,\leftrightharpoons\, 2$). We will, therefore, now focus on how to solve the
interaction terms (we shall find that the self terms are
in fact zero in this calculation).

In order to compute firstly the term of order $1/c$ in (\ref{wij}) to the
1.5PN order, we notice that each of the partial derivatives in (\ref{wijexp2}),
which acts at the field point $\mathbf{x}$, can be transformed into a
derivative acting on the source point, either $\mathbf{y}_{1}$ or
$\mathbf{y}_{2}$, and thus, one can merge the factors together and factor the
differential operator outside the integral. Hence, for instance,
\begin{equation}
\label{spatialderivs}
\mathop{\mathrm{FP}}_{B = 0} \int d^{3}\mathbf{x}
\,\vert\mathbf{x}\vert^{B}\,\partial_i
\left(\frac{1}{r_{1}}\right)\partial_j\left(\frac{1}{r_{2}}\right) =
\frac{\partial}{\partial y_1^i}\frac{\partial}{\partial
y_2^j}\,\mathop{\mathrm{FP}}_{B = 0}\int
d^{3}\mathbf{x}\,\frac{\vert\mathbf{x}\vert^{B}}{r_1r_2}.
\end{equation}
Now, the remaining integral in the R.H.S. of (\ref{spatialderivs}) can be
operationally computed as a particular case of the known elementary integral
called $Y_L$. In fact, we shall need two different types of such integrals,
defined by,
\begin{subequations}\begin{eqnarray}
\label{yl} 
Y_{L} (\mathbf{y}_1, \mathbf{y}_2) & = & \, - \frac{1}{2 \pi} \,
\mathop{\mathrm{FP}}_{B = 0} \int d^{3} \mathbf{x} \, \vert\mathbf{x}\vert^{B}
\, \frac{\hat{x}_{L}}{r_{1}r_{2}},\\
\label{tl}
T_{L} (\mathbf{y}_1, \mathbf{y}_2) & = & \, - \frac{1}{2 \pi} \,
\mathop{\mathrm{FP}}_{B = 0} \int d^{3} \mathbf{x} \, \vert\mathbf{x}\vert^{B} \,
\hat{x}_{L}\, \frac{r_{1}}{r_{2}},
\end{eqnarray}\end{subequations}
where $L\equiv i_1\cdots i_\ell$ is a multi-spatial index of order $\ell$,
$\hat{x}_{L}$ denotes the STF product of spatial vectors:
$\hat{x}_{L}\equiv\mathrm{STF}(x^{i_1}\cdots x^{i_\ell})$ also denoted
$x_{\langle L\rangle}\equiv \hat{x}_{L}$, and the factor $-1/2\pi$ has been
installed for later convenience. These integrals were introduced and computed
in \cite{DI91a,B95,BIJ02}. The general expressions are given in their simplest
form by (see \cite{BIJ02} for a detailed derivation),
\begin{subequations}\label{YLTL}\begin{eqnarray}\label{y12}
Y_{L} & = & \frac{r_{12}}{\ell+1} \,
\sum_{p = 0}^{\ell} \, y_{1}^{\langle L-P} y_{2}^{P \rangle},\\
\label{t12}
T_{L} & = & \frac{r_{12}^3}{3 (\ell+1)(\ell+2)} \, \sum_{p = 0}^{\ell} \, (p +
1) \, y_{1}^{\langle L-P} y_{2}^{P \rangle}.
\end{eqnarray}\end{subequations}
Notice that the finite part operator $\mathop{\mathrm{FP}}_{B = 0}$ plays a
crucial role in such calculations; it removes any divergency at infinity and
makes our computation perfectly clean and well-controlled. From the particular
case $\ell=0$, we find $Y=r_{12}$ and $T=r_{12}^3/6$, and we thus obtain the
integrals needed to compute the $1/c$ term in Eq.~(\ref{wij}).

Consider next the retardation-like $1/c^3$ term in Eq.~(\ref{wij}); as shown
by (\ref{wij}) and (\ref{wijexp2}), this consists in essence of finding the
solution to the integral of the type,
\begin{equation}\label{dev}
\int d^{3}\mathbf{x} \, \vert\mathbf{x}\vert^{B} \,
\frac{\vert\mathbf{x}'-\mathbf{x}\vert^{2}}{r_{1} r_{2}} = \int
d^{3}\mathbf{x}\, \vert\mathbf{x}\vert^{B} \,\frac{\vert\mathbf{x}\vert^2}{r_1
r_2} - 2 \,{x'}^k\int d^{3}\mathbf{x}\, \vert\mathbf{x}\vert^{B}
\,\frac{x^k}{r_1 r_2} + \vert\mathbf{x}'\vert^2 \int
d^{3}\mathbf{x}\,\frac{\vert\mathbf{x}\vert^{B}}{r_1 r_2}.
\end{equation}
As we see, the second and last terms in the R.H.S. can be computed from $Y_L$
with $\ell=1$ and $\ell=0$ respectively. From (\ref{y12}), one obtains $Y_k =
r_{12}\,(y_1^k+y_2^k)/2$. The first term, on the other hand, involves
\textit{a priori} a new structure, and one defines an elementary function for
this term, which reads as,\,\footnote{Actually, one easily checks that this
  function is related to the previous ones by,
$$S_{L} = \left(1-2y_1^k\frac{\partial}{\partial
y_1^k}\right)T_L+\mathbf{y}_1^2\,Y_L.$$}
\begin{equation}\label{SL} 
S_{L} (\mathbf{y}_1, \mathbf{y}_2) = -\frac{1}{2\pi}\mathop{\mathrm{FP}}_{B=0}
\int d^{3} \mathbf{x} \, \vert\mathbf{x}\vert^{B+2}\, \frac{\hat{x}_{L}}{r_{1}
r_{2}}.
\end{equation}
An explicit expression for this elementary function is (see \cite{BIJ02}),
\begin{eqnarray}
\label{sl}
S_{L} (\mathbf{y}_1, \mathbf{y}_2) & = & \frac{r_{12}}{(\ell+1)(\ell+2)} \,
\sum_{p = 0}^{\ell} y_{1}^{\langle L-P} y_{2}^{P \rangle} \nonumber\\
&&\quad\times\left[(\ell + 1 - p) \mathbf{y}_1^{2} - \frac{2}{3}(p+1)(\ell + 1
- p)r_{12}^{2} + (p + 1)\mathbf{y}_{2}^{2}\right],
\end{eqnarray}
which thus permits us to close our computation of the above retardation-like
odd terms. (The self terms, obtained from the interaction terms by the limit
$\mathbf{y}_2 \rightarrow \mathbf{y}_1$, are zero.)

Summarising, we find that our potential $\hat{W}_{ij}^\mathrm{(QNC)}$ admits
the following odd parts. At the 0.5PN level, it is given by a mere function of
time $t$ (through the time dependence of the source points $\mathbf{y}_1$,
$\mathbf{y}_2$ and the velocities $\mathbf{v}_1$, $\mathbf{v}_2$). For later
convenience, we introduce a special notation for it, and pose,
\begin{subequations}\label{wij1}\begin{eqnarray}
\mathop{\hat{W}}_{(1)}{}_{ij}^\mathrm{(QNC)} & = &
\mathop{\alpha}_{(1)}{}_{\!\!ij}(t),\\
\label{alpha1}\mathop{\alpha}_{(1)}{}_{\!\!ij} & \equiv & G^2 m_1
m_2 \frac{d}{dt} \left[\frac{\partial^2\,r_{12}}{\partial y_1^{(i}\partial
y_2^{j)}}\right].
\end{eqnarray}\end{subequations}
At the next 1.5PN order, the potential is given by a quadratic function of the
position, which we express into the form,
\begin{eqnarray}
\label{wij3-NCS-expandedII}
\mathop{\hat{W}}_{(3)}{}_{ij}^\mathrm{(QNC)} & = & \,
\mathop{\alpha}_{(3)}{}_{\!\!ij}(t) + x_k
\,\mathop{\beta}_{(3)}{}_{\!\!ij}^{\!k} (t) + \mathbf{x}^2
\mathop{\gamma}_{(3)}{}_{\!\!ij}(t),
\end{eqnarray}
where the functions of time, as introduced, are respectively given by,
\begin{subequations}\label{alphabetagamma}\begin{eqnarray} 
\nonumber \mathop{\alpha}_{(3)}{}_{\!\!ij} & = & \frac{G^2}{2} \frac{d}{dt}
\left[(m_1\mathop{\widetilde{\mu}}_{(2)}{}_{\!\!2}+m_2\mathop{
\widetilde{\mu}}_{(2)}{}_{\!\!1})\frac{\partial^2\,r_{12}}{\partial
y_1^{i}\partial y_2^{j}}\right]\\ \nonumber & + & \frac{G^2 m_1 m_2}{12}
\,\frac{d}{dt} \left[ a_1^k \frac{\partial^3\,r_{12}^3}{\partial
y_2^{(i}\partial y_1^{j)k}} + v_1^k v_1^l \frac{\partial^4\,r_{12}^3}{\partial
y_2^{(i}\partial y_1^{j)kl}}\right. \\
\label{f-wijcNCS} && \quad \left. + \frac{1}{2} \,\frac{d^2}{dt^2}
\frac{\partial^2}{\partial y_1^{i}\partial y_2^{j}}\left( r_{12}
\,\mathbf{y}_1^2 + r_{12} \,\mathbf{y}_2^2 - \frac{2}{3} r_{12}^3
\, \right) \right] + 1 \,\leftrightharpoons\, 2, \\
\label{g-wijcNCS} \mathop{\beta}_{(3)}{}_{\!\!ij}^{\!k} & = & - 
\frac{G^2 m_1 m_2 }{6} \, \frac{d^3}{dt^3} \frac{\partial^2}{\partial
y_1^{(i}\partial y_2^{j)}}\left( r_{12} \,y_{1}^k + r_{12} \, y_{2}^k \right),\\
\label{h-wijcNCS} \mathop{\gamma}_{(3)}{}_{\!\!ij} & = & \, \frac{G^2 m_1 m_2
}{6} \, \frac{d^3}{dt^3} \frac{\partial^2\,r_{12}}{\partial y_1^{(i}\partial
y_2^{j)}}.
\end{eqnarray}\end{subequations}
As we shall see subsequently, the form of (\ref{wij3-NCS-expandedII}) is of
particular use when one derives the expression of the crucial cubic-order
potential.

\subsection{Cubic non-compact support potentials}\label{secIVc}

In order to illustrate the more difficult types of computation which involve
both Poisson-like and retardation-like spatial integrals, we consider the
example of the part of the non compact support potential $\hat{X}$, referred
to as cubic non-compact (CNC) in Eq.~(\ref{cubpot}), and given by,
\begin{equation}\label{cubpot'}
\hat{X}^{\mathrm{(CNC)}} \equiv
\Box_{\mathcal{R}}^{-1}\left[\hat{W}_{ij}^{\mathrm{(QNC)}}
\partial^2_{ij} V \right],
\end{equation}
in which we recall that $\hat{W}_{ij}^{\mathrm{(QNC)}}$ is defined by
(\ref{Wijdef}). It is necessary to compute the odd term in
$\hat{X}^\mathrm{(CNC)}$ up to the required order 1.5PN. By expanding the
retardations up to this level, we obtain,
\begin{eqnarray}
\nonumber \hat{X}^{\mathrm{(CNC)}} & = & \mathop{\mathrm{FP}}_{B = 0} \left\{
\Delta^{-1} \left[ r^B\,\hat{W}_{ij}^\mathrm{(QNC)} \partial^2_{ij} V \right] +
\frac{1}{c^2} \frac{\partial^2}{\partial t^2}\,\Delta^{-2} \left[
r^B\,\hat{W}_{ij}^\mathrm{(QNC)} \partial^2_{ij} V \right] \right.\nonumber\\ &&
\left. + \frac{1}{4 \pi c}\,\frac{d}{dt} \int d^{3}\mathbf{x}
\,\vert\mathbf{x}\vert^B\, \hat{W}_{ij}^\mathrm{(QNC)} \partial^2_{ij} V +
\frac{1}{24 \pi c^{3}} \,\frac{\partial^3}{\partial t^3} \int
d^{3}\mathbf{x}\,\vert\mathbf{x}\vert^B\,\vert\mathbf{x}'-\mathbf{x}\vert^{2}
\, \hat{W}_{ij}^\mathrm{(QNC)} \partial^2_{ij} V \right\} \nonumber\\ && +
\mathcal{O} \left(\frac{1}{c^{4}}\right).\label{xexp}
\end{eqnarray}
The last two explicit terms in the latter expression represent the
retardation-like odd terms and are computed using the same techniques as in
the case of the quadratic non-compact support potential,
$\hat{W}_{ij}^{\mathrm{(QNC)}}$ (see the previous Section).

In contrast, however, to the case of the QNC potential, we see from
Eq.~(\ref{xexp}) that since the source of the Poisson integral will contain
some odd terms at the $1/c$ and $1/c^3$ levels, there are, in addition to the
retardation-like terms, some ``Poisson-like'' contributions to the odd terms in
the CNC potential [the first two terms in Eq.~(\ref{xexp})]. Since there are
no terms at orders 0.5PN and 1.5PN in the gradient of $V$, one finds that the
0.5PN and 1.5PN pieces in the source term, $\hat{W}_{ij}^\mathrm{(QNC)}
\partial^2_{ij} V$, come only from the odd terms in the potential
$\hat{W}_{ij}^\mathrm{(QNC)}$ itself, which has already been computed in
Eqs.~(\ref{wij1})--(\ref{alphabetagamma}). We show only the more difficult
case of the 1.5PN order. Hence, the source term for the Poisson integral is in
the form,
\begin{equation} \label{poissonsourceodd2}
\left[ \hat{W}_{ij}^\mathrm{(QNC)} \partial^2_{ij} V
\right]_{(3)}=\mathop{\alpha}_{(3)}{}_{\!\!ij}\,\partial^2_{ij}
\mathop{V}_{(0)}+\mathop{\alpha}_{(1)}{}_{\!\!ij}\,\partial^2_{ij}
\mathop{V}_{(2)}+\mathop{\beta}_{(3)}{}_{\!\!ij}^{\!k}\,x^k\,\partial^2_{ij}
\mathop{V}_{(0)}
+\mathop{\gamma}_{(3)}{}_{\!\!ij}\,\mathbf{x}^2\,\partial^2_{ij}
\mathop{V}_{(0)}.
\end{equation}
We work out this expression using the known even part of the $V$ potential. By
thus transforming the derivatives with respect to the field point into
derivatives with respect to the source points (using
$\partial_i=-\partial/\partial y_1^i$ when acting on a function of $r_1$), one
arrives at a new expression which can be immediately integrated, with the
result:
\begin{eqnarray} \label{poissonxodd2}
\mathop{\mathrm{FP}}_{B = 0} \Delta^{-1} \left[
r^B\,\hat{W}_{ij}^\mathrm{(QNC)} \partial^2_{ij} V
\right]_{(3)}&=&\mathop{\alpha}_{(3)}{}_{\!\!ij}\,\frac{\partial^2}{\partial
y_1^{ij}} \left[G m_1 \frac{r_1}{2}\right] +
\mathop{\alpha}_{(1)}{}_{\!\!ij}\,\frac{\partial^2}{\partial y_1^{ij}} \left[G
\mathop{\widetilde{\mu}}_{(2)}{}_{\!\!1}\frac{r_1}{2}+\frac{G
m_1}{24}\frac{\partial^2r_1^3}{\partial t^2}\right]\nonumber\\
&+&\mathop{\beta}_{(3)}{}_{\!\!ij}^{\!k}\,\frac{\partial^2}{\partial y_1^{ij}}
\left[\frac{G m_1}{4}\left(x^k+y^k_1\right) r_1\right] \nonumber\\
&+&\mathop{\gamma}_{(3)}{}_{\!\!ij}\,\frac{\partial^2}{\partial y_1^{ij}}
\left[G m_1\,\left(\frac{r_1^3}{12}+\frac{1}{2}y_1^k x^k r_1\right)\right]+
1\,\leftrightharpoons\, 2,
\end{eqnarray}
where we recall that the time-dependent coefficients have been given in
Eqs.~(\ref{alpha1}) and (\ref{alphabetagamma}). Concerning the other
Poisson-like integral, the computation is easier because it is required only
at 0.5PN order, and we simply obtain,
\begin{eqnarray} \label{poissonxodd5}
\mathop{\mathrm{FP}}_{B = 0} \Delta^{-2} \left[
r^B\,\hat{W}_{ij}^\mathrm{(QNC)} \partial^2_{ij} V
\right]_{(1)}&=&\mathop{\alpha}_{(1)}{}_{\!\!ij}\,\frac{\partial^2}{\partial
y_1^{ij}} \left[G m_1 \frac{r_1^3}{24}\right] + 1\,\leftrightharpoons\, 2.
\end{eqnarray}
Having in hand all the required terms at any field point $\mathbf{x}$, we can
now compute the appropriate gradients of these potentials which enter the
equations of motion (\ref{eqnmotion1})--(\ref{eqnmotion2}), and then obtain
their value at the location of the particle 1 using Hadamard's partie finie
(\ref{F1}). For instance, the gradient of the term (\ref{poissonxodd5}), taken
at the location of the particle 1, reads,
\begin{equation}\label{poissonxoddII}
\left(\partial_k \mathop{\mathrm{FP}}_{B = 0}\Delta^{-2} \left[
r^B\,\hat{W}_{ij}^\mathrm{(QNC)} \partial^2_{ij} V \right]_{(1)}\right)_1 =
\mathop{\alpha}_{(1)}{}_{\!\!ij}\,\frac{\partial^3}{\partial y_1^{ijk}}
\left[G m_2 \frac{r_{12}^3}{24}\right],
\end{equation}
where $r_{12}=\vert\mathbf{y}_1-\mathbf{y}_2\vert$ represents the particles'
separation.

\subsection{Alternative derivation of the contribution of NC potentials}\label{secIVd}

As Sections \ref{secIVb} and \ref{secIVc} discuss, the previous method
consisted of first computing all the required NC potentials \textit{at any
field point} $\mathbf{x}$, and then accounting for their contributions to the
equations of motion of particle 1 (say), by applying the Hadamard partie
finie (\ref{F1}). One can, however, also compute \textit{directly} the value
at point 1 of any potential given in the form of a Poisson integral. This
alternative technique is indeed used in the computation of the equations of
motion at 3PN order in Ref.~\cite{BFeom}. In the problem of the 3PN equations
of motion, it was impossible to obtain the value of all the potentials at an
arbitrary field point $\mathbf{x}$, and thus, the complete result could only
be achieved by using the latter technique. In the present case of the 3.5PN
term, as seen previously, all the potentials at any field point could be
computed by analytical methods. We are, therefore, able to perform an
important verification of the result by directly evaluating the potentials
``on the line'', \textit{i.e.} at the source point $\mathbf{y}_1$ (say).

Such a direct evaluation on the line applies equally well to the computation
of retardation-like odd terms or the Poisson and Poisson-like integrals. It
proceeds from the Hadamard partie-finie formalism reviewed in Section
\ref{secIII}, and, in particular, from the expression of the partie finie
integral (\ref{PfF}). We check that all the retardation-like odd terms can be
cast in the form of some sum of partie finie integrals of the type
(\ref{PfF}), which are eventually multiplied by some spatial vector positions;
this is due to the factor, say $\vert\mathbf{x}-\mathbf{x}'\vert^{2k}$, which
enters such integrals, and can always be expanded as in Eq.~(\ref{dev}). The
computation using this technique of the Poisson-like terms is more complicated
but the full proofs are given in section \ref{secV} of Ref.~\cite{BFreg}.
Here, we provide only a summary of this computation, which has been
systematicall performed to obtain full confirmation of our result. Typically,
one is dealing with the Poisson integral of some $F\in\mathcal{F}$, say,
\begin{equation}
\label{intP}
P (\mathbf{x}') = -\frac{1}{4\pi} \,\mathrm{Pf} \, \int \frac{d^3
\mathbf{x}}{\vert\mathbf{x}-\mathbf{x}'\vert} \, F(\mathbf{x}),
\end{equation}
where the symbol Pf refers to Hadamard's partie finie in the sense of
Eq.~(\ref{PfF}). We are interested in the value of this Poisson integral when
$\mathbf{x}'$ tends to the particle's position $\mathbf{y}_1$, and this value
takes the meaning of Eq.~(\ref{F1}). In other words, the computation of the
quantity $(P)_1$, and also of the corresponding gradient, $(\partial_i P)_1$,
are required. They have already been given in Eqs.~(5.5) and (5.17a) of
\cite{BFreg}, and read as,
\begin{subequations}\label{pHad}\begin{eqnarray}
\label{pHad1}
(P)_1 & = & -\frac{1}{4\pi}\,\mathrm{Pf} \int d^3
\mathbf{x}\,\frac{F}{r_1} + \left[ \ln \left(\frac{r_{1}'}{s_1}\right)
- 1 \right] \, (r_1^2 F)_1,\\
\label{dipHad1}
(\partial_i P)_1 & = & -\frac{1}{4\pi}\,\mathrm{Pf} \int d^3
\mathbf{x}\,\frac{n_1^i}{r_1^2} F + \ln \left(\frac{r_{1}'}{s_1}\right) \,
(n_1^i r_1 F)_1.
\end{eqnarray}\end{subequations}
Recall that the partie finie integral (\ref{PfF}) depends on two constants
$s_1$ and $s_2$. In addition, $r'_1=\vert\mathbf{x}'-\mathbf{y}_1\vert$ in
(\ref{pHad}) is a ``constant'' which tends toward zero when evaluating the
partie finie. Although these constants played an important role in the
computation of the equations of motion at 3PN order, none of them appear in
the present work. For this calculation, we also require the formulas
concerning the twice-iterated Poisson integral, and the results were provided
by the equations (5.16) and (5.17b) in \cite{BFreg}.

We have found by using the above formulas that the direct method is in
complete agreement for all the terms with the analytical methods reviewed in
Sections \ref{secIVb} and \ref{secIVc}. However, notice that for the agreement
to work, one must crucially take into account, in addition to the formulas
such as (\ref{pHad}), the contribution of the distributional part of
derivatives. For this calculation, we use formula (\ref{distderiv1}),
which gives the distributional derivative in the extended Hadamard
regularisation.

\section{The 3.5PN compact binary acceleration}\label{secV}

We finally present our result, which gives the complete radiation reaction
force in the equations of motion of the compact binary at the 3.5PN order, for
general orbits and in a general harmonic coordinate frame. We write the 3.5PN
acceleration of the particle 1, say, in the form,
\begin{equation}\label{acc1}
\mathbf{a}_1 = \mathbf{A}_{1}^\mathrm{N} +
\frac{1}{c^2}\mathbf{A}_{1}^\mathrm{1PN} +
\frac{1}{c^4}\mathbf{A}_{1}^\mathrm{2PN} +
\frac{1}{c^5}\mathbf{A}_{1}^\mathrm{2.5PN} +
\frac{1}{c^6}\mathbf{A}_{1}^\mathrm{3PN} +
\frac{1}{c^7}\mathbf{A}_{1}^\mathrm{3.5PN} +
\mathcal{O}\left(\frac{1}{c^8}\right),
\end{equation}
where the first term is given by the famous Newtonian law,
\begin{equation}\label{accN}
\mathbf{A}_{1}^\mathrm{N} = -\frac{G m_2}{r_{12}^2}\,\mathbf{n}_{12}.
\end{equation}
The acceleration of the second particle is obtained by exchanging all the
labels $1\,\leftrightharpoons\, 2$. The conservative (with even-parity)
approximations 1PN, 2PN and 3PN have been computed elsewhere; they are
thoroughly given by Eq.~(7.16) in \cite{BFeom} or by Eq.~(131) in
\cite{Bliving}, together with the value of the ambiguity parameter
$\lambda=-1987/3080$ computed in \cite{BDE04}.

The result central to our study concerns the radiation reaction (odd-order)
acceleration coefficients $\mathbf{A}_{1}^\mathrm{2.5PN}$ and
$\mathbf{A}_{1}^\mathrm{3.5PN}$. We find, for the 2.5PN term,\,\footnote{Our
notation is $r_{12}=\vert\mathbf{y}_1-\mathbf{y}_2\vert$,
$\mathbf{n}_{12}=(\mathbf{y}_1-\mathbf{y}_2)/r_{12}$, and
$\mathbf{v}_1=d\mathbf{y}_1/dt$, $\mathbf{a}_1=d\mathbf{v}_1/dt$ for the
harmonic-coordinates velocity and acceleration (together with
$1\,\leftrightharpoons\, 2$). We pose
$\mathbf{v}_{12}=\mathbf{v}_1-\mathbf{v}_2$ for the relative velocity. The
parenthesis indicate the usual Euclidean scalar product, \textit{e.g.} $(n_{12}
v_{12})=\mathbf{n}_{12}\cdot\mathbf{v}_{12}$.}
\begin{eqnarray}
\mathbf{A}_{1}^\mathrm{2.5PN} & = & \frac{4 \,G^2 m_1 m_2}{5 \,r_{12}^3}
\left( (n_{12} v_{12}) \left[ - 6 \,\frac{G m_1}{r_{12}} + \frac{52}{3}
\,\frac{G m_2}{r_{12}} + 3 \,v_{12}^2 \right]\mathbf{n}_{12} \right.
\nonumber\\ & &\qquad\qquad\quad \left. + \left[ 2 \, \frac{G m_1}{r_{12}} - 8
\, \frac{G m_2}{r_{12}} - v_{12}^2 \right]\mathbf{v}_{12} \right).
\label{a1_5}
\end{eqnarray}
This result is in perfect agreement with previous calculations in a general
harmonic coordinate frame
\cite{DD81a,D82,D83houches,BFP98,IFA01,BFeom,itoh1,itoh2}, as well as in the
center-of-mass frame \cite{PW02}. The new result is the complete expression of
the 3.5PN coefficient in an arbitrary frame which we find to be given by,
\allowdisplaybreaks{\begin{eqnarray} \mathbf{A}_{1}^\mathrm{3.5PN} &=&
\frac{G^2 m_1 m_2}{r_{12}^3} \, \left\{ \frac{G^2 m_1^2}{r_{12}^2} \, \left[
\left( \, \frac{3992}{105} (n_{12} v_1) - \frac{4328}{105} (n_{12} v_2)
\right)\mathbf{n}_{12} - \frac{184}{21} \mathbf{v}_{12} \, \right]
\right.\nonumber\\
&& + \frac{G^2 m_1 m_2}{r_{12}^3}
\, \left[ \left( \, - \frac{13576}{105} (n_{12} v_1) \, +
\, \frac{2872}{21} (n_{12} v_2) \, \right)\mathbf{n}_{12} + \frac{6224}{105}
\mathbf{v}_{12} \, \right] \nonumber\\ 
&& + \frac{G^2 m_2^2}{r_{12}^3} \, \left[ - \frac{3172}{21} (n_{12}
v_{12})\,\mathbf{n}_{12} + \frac{6388}{105} \mathbf{v}_{12} \right]
\nonumber\\
&& + \frac{G m_1}{r_{12}} \, \left[ \left( \, 48 (n_{12} v_1)^3 - \frac{696}{5}
(n_{12} v_1)^2 (n_{12} v_2) + \frac{744}{5} (n_{12} v_1) (n_{12} v_2)^2
\right.\right.\nonumber\\ &&\quad\qquad\qquad \left.\left. - \frac{288}{5}
(n_{12} v_2)^3 -\frac{4888}{105} (n_{12} v_1) v_1^2 + \frac{5056}{105}
(n_{12} v_2) v_1^2 \right.\right.\nonumber\\ &&\quad\qquad\qquad
\left.\left. + \frac{2056}{21} (n_{12} v_1) (v_1 v_2) - \frac{2224}{21}
(n_{12} v_2) (v_1 v_2) \right.\right.\nonumber\\ &&\quad\qquad\qquad
\left.\left. - \frac{1028}{21} (n_{12} v_1) v_2^2 + \frac{5812}{105}
(n_{12} v_2) v_2^2 \right)\mathbf{n}_{12} \right. \nonumber\\
&&\quad\qquad \left. + \left( \frac{52}{15} (n_{12} v_1)^2 - \frac{56}{15}
(n_{12} v_1) (n_{12} v_2) - \frac{44}{15} (n_{12} v_2)^2 - \frac{132}{35}
v_1^2 \right.\right. \nonumber\\ &&\quad\qquad\qquad \left.\left. +
\frac{152}{35} (v_1 v_2) - \frac{48}{35} v_2^2 \right)\mathbf{v}_{12} \right]
\nonumber\\
&& + \frac{G m_2}{r_{12}} \, \left[ \left( - \frac{582}{5} (n_{12} v_1)^3 +
\frac{1746}{5} (n_{12} v_1)^2 (n_{12} v_2) - \frac{1954}{5} (n_{12} v_1)
(n_{12} v_2)^2 \right.\right.\nonumber\\ &&\quad\qquad\qquad
\left.\left. + \, 158 (n_{12} v_2)^3 +\frac{3568}{105} (n_{12} v_{12}) (v_1
v_1) - \frac{2864}{35} (n_{12} v_1) (v_1 v_2) \right.\right.\nonumber\\
&&\quad\qquad\qquad \left.\left. + \frac{10048}{105} (n_{12} v_2) (v_1
v_2) +\frac{1432}{35} (n_{12} v_1) v_2^2 
- \frac{5752}{105} (n_{12} v_2) v_2^2 \right)\mathbf{n}_{12}\right. \nonumber\\
&&\quad\qquad \left. + \left( \frac{454}{15} (n_{12} v_1)^2 - \frac{372}{5}
(n_{12} v_1) (n_{12} v_2) + \frac{854}{15} (n_{12} v_2)^2 - \frac{152}{21}
v_1^2 \right.\right.\nonumber\\ &&\quad\qquad\qquad \left.\left. +
\frac{2864}{105} (v_1 v_2) - \frac{1768}{105} v_2^2 \right)\mathbf{v}_{12}
\right]\nonumber\\
&& + \left( - 56 (n_{12} v_{12})^5 + 60 (n_{12} v_{1})^3 v_{12}^2 - 180
(n_{12} v_{1})^2 (n_{12} v_{2}) v_{12}^2 + 174 (n_{12} v_{1}) (n_{12} v_{2})^2
v_{12}^2 \right. \nonumber\\ &&\quad\qquad\qquad \left.\left. - 54 (n_{12}
v_{2})^3 v_{12}^2 -\frac{246}{35} (n_{12} v_{12}) v_1^4 + \frac{1068}{35}
(n_{12} v_1) v_1^2 (v_1 v_2) \right.\right. \nonumber\\ &&\quad\qquad\qquad
\left.\left. - \frac{984}{35} (n_{12} v_2) v_1^2 (v_1 v_2) - \frac{1068}{35}
(n_{12} v_1) (v_1 v_2)^2 + \frac{180}{7} (n_{12} v_2) (v_1 v_2)^2
\right.\right. \nonumber\\ &&\quad\qquad\qquad \left.\left. - \frac{534}{35}
(n_{12} v_1) v_1^2 v_2^2 + \frac{90}{7} (n_{12} v_2) v_1^2 v_2^2 +
\frac{984}{35} (n_{12} v_1) (v_1 v_2) v_2^2 \right.\right. \nonumber\\
&&\quad\qquad\qquad \left.\left. - \frac{732}{35} (n_{12} v_2) (v_1 v_2) v_2^2
- \frac{204}{35} (n_{12} v_1) v_2^4 + \frac{24}{7} (n_{12} v_2) v_2^4
\right)\mathbf{n}_{12}\right. \nonumber\\
&& + \left. \left( 60 (n_{12} v_{12})^4 - \frac{348}{5} (n_{12} v_1)^2
v_{12}^2 + \frac{684}{5} (n_{12} v_1) (n_{12} v_2) v_{12}^2 - 66 (n_{12}
v_2)^2 v_{12}^2 \right.\right. \nonumber\\ &&\quad\qquad\qquad \left.\left. +
\frac{334}{35} v_1^4 - \frac{1336}{35} v_1^2 (v_1 v_2) + \frac{1308}{35} (v_1
v_2)^2 + \frac{654}{35} v_1^2 v_2^2 \right.\right. \nonumber\\
&&\quad\qquad\qquad \left.\left. - \frac{1252}{35} (v_1 v_2) v_2^2 +
\frac{292}{35} v_2^4 \right)\mathbf{v}_{12} \right\}.\label{a1_7}
\end{eqnarray}}\noindent
Recall that we obtain the results (\ref{a1_5}) and (\ref{a1_7}) directly from
(\ref{eqnmotion1}) and (\ref{eqnmotion2}) by summing up all the contributions
of the regularised values of the potentials. The latter were computed
following the two (rather independent) methods proposed in Section
\ref{secIV}.

We next give the result for the 2.5PN and 3.5PN relative accelerations in the
center-of-mass frame. For this calculation, we require the transformation
equations for converting from the positions and velocities of the particles in
the general frame to those in the center-of-mass frame. Naturally, the center
of mass is defined by the nullity of the binary's mass dipole moment at the
required PN order. All these relations have been worked out at the 3PN order
in Ref.~\cite{BI03CM}. Specifically, the transformation equations take the
form,\,\footnote{For center-of-mass quantities, we denote by $\mathbf{x} =
\mathbf{y}_1 - \mathbf{y}_2$, $r=\vert\mathbf{x}\vert$, $\mathbf{n} =
\mathbf{x}/r$ the relative binary's separation (formerly denoted $r_{12}=r$
and $\mathbf{n}_{12}=\mathbf{n}$), and by $\mathbf{v} = d\mathbf{x}/dt =
\mathbf{v}_1 - \mathbf{v}_2$ the relative velocity. The mass parameters are
given as,
$$m=m_1+m_2, \qquad X_1=\frac{m_1}{m}, \qquad X_2=\frac{m_2}{m},
\qquad\nu=X_1X_2.$$}
\begin{subequations}\label{y1y2}\begin{eqnarray}
\mathbf{y}_1 & = & \left[ X_{2} + \nu ( X_1 - X_2 ) \mathcal{P}
\right] \,\mathbf{x} + \nu (X_1-X_2) \mathcal{Q} \,\mathbf{v}, \\
\mathbf{y}_2 & = & \left[ -X_{1} + \nu ( X_1 - X_2 ) \mathcal{P}
\right] \,\mathbf{x} + \nu (X_1-X_2) \mathcal{Q} \,\mathbf{v}.
\end{eqnarray}\end{subequations}
The coefficients $\mathcal{P}$ and $\mathcal{Q}$ are given at 3PN order by
Eqs.~(3.11)--(3.12) in \cite{BI03CM}. For our purposes, we require the
relative 1PN even correction term in $\mathcal{P}$ and $\mathcal{Q}$, since we
are computing the 3.5PN radiation reaction force, which is at relative 1PN
order. In addition, we have found that the 2.5PN odd correction term in the
latter transformation equations, gives a crucial contribution at the 3.5PN
order in the radiation reaction force. Inspection of Eqs.~(3.11)--(3.12) in
\cite{BI03CM}, reveals that the 1PN correction term exists only in the
coefficient $\mathcal{P}$ [since $\mathcal{Q}$ begins at the $1/c^4$ level],
whilst the odd contribution at order $1/c^5$ is proportional to the velocity,
and hence is contained only in $\mathcal{Q}$. Using our previous notation
(\ref{PNF}) for PN coefficients, we have the required terms,
\begin{subequations}\label{P2Q5}\begin{eqnarray}
\mathop{\mathcal{P}}_{(2)} &=& \frac{v^2}{2}-\frac{G\,m}{2 r},\\
\mathop{\mathcal{Q}}_{(5)} &=& \frac{4 G\,m}{5}\left[ v^2 -\frac{2
G\,m}{r}\right].
\end{eqnarray}\end{subequations}
We have inserted Eqs.~(\ref{y1y2})--(\ref{P2Q5}) into our general-frame
result, in order, therefore, to obtain the 3.5PN relative center-of-mass
acceleration in the form,
\begin{eqnarray}\label{acc}
\mathbf{a}&\equiv& \mathbf{a}_1-\mathbf{a}_2 \nonumber\\ &=&
\mathbf{A}^\mathrm{N} + \frac{1}{c^2}\mathbf{A}^\mathrm{1PN} +
\frac{1}{c^4}\mathbf{A}^\mathrm{2PN} + \frac{1}{c^5}\mathbf{A}^\mathrm{2.5PN}
+ \frac{1}{c^6}\mathbf{A}^\mathrm{3PN} +
\frac{1}{c^7}\mathbf{A}^\mathrm{3.5PN} + \mathcal{O}\left(\frac{1}{c^8}\right).
\end{eqnarray}
All the terms up to 3PN order have already been computed in
Eqs.~(3.16)--(3.18) of \cite{BI03CM}. The radiation reaction 2.5PN term reads
as,
\begin{eqnarray}\label{a5cofm}
\mathbf{A}^\mathrm{2.5PN} & = & \frac{8 G^2\,m^2\,\nu}{5r^3}\left\{\left[-
3\frac{G m}{r} - v^2\right]\mathbf{v}+(nv)\left[\frac{17}{3} \frac{G m}{r} + 3
v^2\right]\mathbf{n}\right\},
\end{eqnarray}
whilst the 3.5PN contribution is given as,
\begin{eqnarray}
\mathbf{A}^\mathrm{3.5PN} & = & \frac{G^2 m^2 \nu}{r^3}\left\{\left[\frac{G^2
m^2}{r^2} \, \left( \frac{1060}{21} + \frac{104}{5}\nu \right) + \frac{G
m v^2}{r} \, \left( - \frac{164}{21} - \frac{148}{5} \nu \right)
\right.\right. \nonumber\\ & & \qquad\qquad\left.\left. + v^4\, \left(
\frac{626}{35} + \frac{12}{5} \nu \right) + \frac{G m (nv)^2}{r} \,
\left( \frac{82}{3} + \frac{848}{15} \nu \right) \right.\right.\nonumber\\ & &
\qquad\qquad\left.\left. + v^2 (nv)^2 \left( - \frac{678}{5} - \frac{12}{5}
\nu \right) + 120 (nv)^4 \right]\mathbf{v} \right. \nonumber\\ & & \qquad
\left. + (nv)\left[\frac{G^2 m^2}{r^2} \, \left( - \frac{3956}{35} -
\frac{184}{5} \nu \right) + \frac{G m \,v^2}{r} \, \left( - \frac{692}{35} +
\frac{724}{15} \nu \right)\right.\right. \nonumber\\ & &
\qquad\qquad\left.\left. + v^4 \, \left( - \frac{366}{35} - 12 \nu \right) +
\frac{G m \,(nv)^2}{r} \, \left( - \frac{294}{5} - \frac{376}{5} \nu \right)
\right.\right.\nonumber\\ & & \qquad\qquad\left.\left.+ v^2 (nv)^2 \, \left(
114 + 12 \nu \right) - 112 (nv)^4 \right] \mathbf{n} \right\}.\label{a7cofm}
\end{eqnarray}

The expressions (\ref{a5cofm}) and (\ref{a7cofm}) allow for an important and
mutual consistency check with the results of Iyer and Will \cite{IW93,IW95}.
These authors computed the radiation reaction force at 3.5PN order for compact
binary systems in a class of coordinate systems, by applying the energy and
angular momentum balance equations to relative 1PN order. As a consequence of
the gauge freedom, the radiation reaction at 2.5PN order depends on two gauge
parameters, denoted $\alpha$ and $\beta$, whilst at the 3.5PN order, it
depends on six further gauge parameters, denoted $\delta_A$, $A=1,\cdots,6$
\cite{IW93,IW95}. The eight parameters, $\alpha,\,\beta,\,\delta_A$, were
proved to correspond exactly to the unconstrained degrees of freedom which
relate to coordinate transformations. For instance, the 3.5PN radiation
reaction potentials of Ref.~\cite{B97} were shown when evaluated in the case
of compact binaries to correspond to a unique, self-consistent choice of all
these gauge parameters \cite{IW93,IW95}.

In the present case, one also finds perfect agreement between our
specific expressions, (\ref{a5cofm}) and (\ref{a7cofm}), derived from ``first
principles'', with the end result of Ref.~\cite{IW93,IW95}. This is provided
that the eight parameters, $\alpha,\,\beta,\,\delta_A$, assume some constant
values reflecting the present choice of harmonic coordinates. By comparing our
results (\ref{a5cofm}) and (\ref{a7cofm}) with Eqs.~(2.12) together with
(2.18) in \cite{IW95} (notice the change of notation:
$\varepsilon_5\rightarrow\delta_6$), we indeed obtain a unique and consistent
choice for these parameters, given by,
\begin{center}\begin{equation}
\begin{tabular}{p{0.8cm}cp{4.0cm}p{0.8cm}cp{4.0cm}}
$\alpha$ & = & $~-1$, & $\beta$ & = & $~0$,\\ \\ $\delta_1$& = &
$~\frac{271}{28} + 6 \nu$, & $\delta_2$ & = & $~- \, \frac{77}{4} \, - \,
\frac{3}{2} \nu$,\\ \\ $\delta_3$& = & $~\frac{79}{14} - \frac{92}{7} \nu$, &
$\delta_4$ & = & $~10$,\\ \\ $\delta_5$ & = & $~\frac{5}{42} + \frac{242}{21}
\nu$, & $\delta_{6}$& = & $~- \frac{439}{28} \, + \frac{18}{7} \nu$.\\
\end{tabular} 
\label{resultparameters}\end{equation}\end{center}
These values correspond to harmonic coordinates. For these, we find complete
agreement with the result obtained (also from first principles) by Pati and
Will \cite{PW02} (see also \cite{MW03}). On the other hand, some other values
correspond to ADM coordinates, as computed by K{\"o}nigsd{\"o}rffer \textit{et
al.} \cite{KFS03}. Still other values for these parameters, obtained in
\cite{IW95}, correspond to the 3.5PN radiation reaction potentials \cite{B97}
valid in extended Burke-Thorne gauge.

In conclusion, we have computed at 3.5PN order the radiation reaction effect
in the local equations of motion of a compact binary system in a general
harmonic coordinates frame. The result was derived using a direct PN iterated
expansion of the metric in harmonic coordinates at the 3.5PN approximation
derived in Refs.~\cite{BFP98,BFeom}. The 3.5PN metric is expressed as a
function of a particular set of non-linear retarded potentials, defined for
any general smooth ``hydrodynamical'' matter distribution. The existence of
singular functions and divergent integrals, a consequence of our modelisation
by two delta-function singularities, required the Hadamard partie finie
regularisation in order to remove each particle's infinite self-field.
Analytical techniques were used to solve the more complicated integrals of the
NC supported distribution of the gravitational field, resulting in a final
expression in closed analytic form. We found that the 3.5PN term in the
equations of motion in the center-of-mass frame is perfectly consistent with
the general expression of the 1PN radiation reaction force, derived by energy
and angular-momentum balance equations in a class of coordinate systems
\cite{IW93,IW95}. This study thus confirms that all the results to date for
the 3.5PN equations of motion, computed either by ``direct'' iterative PN
expansion or by ``indirect'' energy and angular momentum balance
considerations, are fully in agreement with each other. Since the equations of
motion up to the 3PN order have already been derived elsewhere
(\cite{BF00,BFeom,BDE04,itoh1,itoh2} in harmonic coordinates,
\cite{JaraS98,JaraS99,DJSdim} in ADM coordinates), we conclude that the
problem of the local equations of motion of compact binaries is solved up to
3.5PN order.

\acknowledgments

One of us (S.N.) would like to thank The Leverhulme Trust Study Abroad Program
and the Entente Cordiale Scholarship for financial support.

\bibliography{NB04}

\end{document}